\documentclass[sigconf]{acmart}
\AtBeginDocument{%
  }

\usepackage{comment}
\usepackage{algorithm}
\usepackage{algorithmic}
\usepackage{amsmath}
\usepackage{amsthm}
\usepackage{bm}
\usepackage{mdframed}

\newtheorem{definition}{Definition}
\newtheorem{proposition}{Proposition}
\newtheorem{remark}{Remark}
\newtheorem{assumption}{Assumption}
\newtheorem{theorem}{Theorem}
\newtheorem{lemma}{Lemma}

\newcommand{\bs}{\boldsymbol}

\newcommand{\fb}[1]{\textcolor{red}{FB: #1}}
\newcommand{\zz}[1]{\textcolor{magenta}{ZZ: #1}}

\setcopyright{acmlicensed}
\copyrightyear{2018}
\acmYear{2018}
\acmDOI{XXXXXXX.XXXXXXX}

\acmConference[Conference acronym 'XX]{Make sure to enter the correct
conference title from your rights confirmation email}{June 03--05,
  2018}{Woodstock, NY}

\acmISBN{978-1-4503-XXXX-X/2018/06}

\begin{document}

\title{Geometry-Evolving Barrier Functions for Safety Evaluation in Dynamic Environments}

\author{Zhi Zhang}
\email{zhi.zhang90@outlook.com}
\affiliation{%
  \institution{Imperial College London}
  \city{London}
  \country{UK}}
\author{Francesco Belardinelli}
\email{francesco.belardinelli@imperial.ac.uk}
\affiliation{%
  \institution{Imperial College London}
  \city{London}
  \country{UK}}

\author{Alessandro Abate}
\email{alessandro.abate@cs.ox.ac.uk}
\affiliation{%
  \institution{University of Oxford}
  \city{Oxford}
  \country{UK}}

\renewcommand{\shortauthors}{Zhang et al.}

\begin{abstract}
Autonomous systems operating in dynamic environments often involve multiple interacting agents, giving rise to high-dimensional safety evaluation problems. Hamilton--Jacobi (HJ) reachability provides rigorous finite-horizon safety guarantees, but
solving a global HJ partial differential equation (PDE) over the entire state space leads to prohibitive computational complexity.
In this paper, we introduce Geometry-Evolving Barrier Functions (GEBFs), a novel representation for dynamic safety evaluation that propagates local safety geometry of the solution to a terminal-value HJ PDE.
By evolving the value, gradient, and curvature information along a Hamiltonian-induced reference trajectory, GEBFs provide a compact representation of the local HJ safety landscape.
Under suitable local regularity assumptions, we establish consistency between the GEBF and the value, gradient, and Hessian of the HJ solution along an HJ-optimal characteristic.
Building on these theoretical results, we develop an iterative algorithm that couples local geometric propagation with trajectory refinement.
We further establish a trajectory-wise one-sided relationship between the GEBF and HJ value function, together with a fixed-point consistency result for the iterative scheme.
Experiments on widely adopted agent--obstacle collision-avoidance benchmarks, across low- and high-dimensional systems, demonstrate that GEBFs accurately capture the evolution of dynamic safety boundaries, while achieving substantially lower computational cost than conventional HJ reachability.
\end{abstract}

\ccsdesc[500]{Do Not Use This Code~Generate the Correct Terms for Your Paper}
\ccsdesc[300]{Do Not Use This Code~Generate the Correct Terms for Your Paper}
\ccsdesc{Do Not Use This Code~Generate the Correct Terms for Your Paper}
\ccsdesc[100]{Do Not Use This Code~Generate the Correct Terms for Your Paper}

\keywords{Hamilton--Jacobi Reachability,
Geometry-Evolving Barrier Functions,
Safety Evaluation,
Dynamic Environments,
Nonlinear Dynamical Systems}

\received{20 February 2007}
\received[revised]{12 March 2009}
\received[accepted]{5 June 2009}

\maketitle

\section{Introduction}

Safety is a fundamental requirement for autonomous systems operating in dynamic environments \cite{amodei2016concrete,hsu2023safety}, including autonomous driving \cite{paden2016survey}, robotics \cite{ma2024conformal,luo2023towards}, and multi-agent systems \cite{garg2024learning}. Unlike static environments, where unsafe regions remain fixed, dynamic environments continuously reshape the boundary between safe and unsafe states through the evolution of both the agent and its surroundings.
Dynamic environments naturally lead to high-dimensional state spaces, making rigorous safety analysis significantly more challenging.
Safe decision-making requires not only identifying whether a state is safe, but also characterizing how the safety boundary evolves over time.
Capturing this evolving safety boundary is therefore central to both safety verification and safety-critical control.

Hamilton–Jacobi (HJ) reachability offers a rigorous mathematical framework for characterizing dynamic safety boundaries by propagating the safety value function backward in time, rather than treating safety as a static property \cite{chen2018hamilton,ganai2024hamilton}.
By solving HJ partial differential equations (PDEs), including their variational inequality formulations, conventional grid-based reachability methods construct global representations of safety value functions over the state space, providing rigorous safety characterizations and associated optimal controls \cite{mitchell2005toolbox,osher2004level}.
However, conventional grid-based HJ computation requires solving these equations over a discretized state space, leading to computational and memory costs that grow exponentially with the system dimension \cite{mitchell2005time,darbon2016algorithms}. Thus, global grid-based HJ reachability becomes computationally prohibitive in high-dimensional systems.


\textcolor{black}{To alleviate this computational bottleneck, scalable HJ methods have been developed through system decomposition, neural approximations, and numerical acceleration techniques~\cite{chen2017exact,bansal2021deepreach,yang2025scalable,herbert2021scalable,bansal2017hamilton}. Beyond global state-space approximations, characteristic- and Hopf-based methods enable pointwise HJ evaluation without global discretization~\cite{chow2019algorithm,yegorov2021perspectives}, while conservative linear-envelope methods provide scalable reachability approximations~\cite{sharpless2026conservative}.
A complementary question, which we pursue here, is whether the local
differential geometry of HJ value function can itself be represented
and propagated through a finite-dimensional function along a trajectory.
Such a representation would provide an explicit description of how local safety geometry inherited from the terminal condition is transported and deformed by system dynamics, including its first- and second-order geometric information.}

\textcolor{black}{Barrier functions provide a natural representation for this purpose:
they encode safety through an explicit scalar function and can be directly incorporated into forward-invariance conditions \cite{ames2019control,zhang2023exact}. Conventional barrier constructions, however, typically begin with a prescribed or separately synthesized barrier geometry and certify its forward invariance, rather than deriving the evolution of that geometry from the HJ value function.
This motivates the question of whether an evolving barrier representation can be constructed directly from the
local differential geometry of the HJ value function, while avoiding the construction of a global HJ solution.}

\paragraph{\bf Contribution.}
In this paper, we introduce {\em Geometry-Evolving Barrier Functions} (GEBFs), a dynamic barrier representation that explicitly captures the evolution of safety boundaries in dynamic environments.
Rather than constructing or approximating a global HJ safety representation over the entire state space, GEBFs target the
dynamical propagation governed by the HJ PDE and propagate the corresponding local safety geometry along a reference trajectory through a compact parametric representation.
In particular, GEBFs capture the evolution of the first- and
second-order geometric information of the HJ value function along Hamiltonian-optimal trajectories, which characterizes the local evolution of the safety boundary under the system dynamics.
Such geometric information provides an interpretable characterization of how safety evolves locally under the system dynamics, revealing not only the safety level but also the local orientation, deformation, and sensitivity of the evolving safety boundary. 
These properties make GEBFs a natural representation for local safety analysis and provide a foundation for future robustness-aware control and motion planning.
By propagating local safety geometry, GEBFs provide an efficient and scalable framework for dynamic safety evaluation in high-dimensional systems, \textcolor{black}{trading global coverage for substantially improved computational efficiency.}

Our main contributions are summarized as follows:
\begin{itemize}
\item We introduce GEBFs, a dynamic quadratic barrier representation that propagates local safety geometry.
We establish their theoretical connection to HJ reachability by proving the consistency of their zeroth- (value), first- (gradient), and second-order (Hessian) information with the terminal-value HJ PDE solution along an HJ-optimal characteristic, under the stated regularity assumptions.
\item We develop an efficient iterative construction algorithm.
The algorithm alternates between propagating a local quadratic safety representation along the current reference trajectory and refining the trajectory using the Hamiltonian-maximizing control induced by the resulting GEBF gradient.
We establish that a trajectory induced by an HJ-optimal control is a fixed point of this geometry–trajectory update, and further establish a trajectory-wise one-sided relation between the GEBF and HJ value function, showing that their discrepancy along a GEBF-induced reference trajectory.

\item We evaluate GEBFs on representative agent–obstacle collision-avoidance benchmarks across low- and high-dimensional systems, demonstrating accurate local approximation of HJ safety geometry with substantially reduced computational cost compared with grid-based HJ PDE computation.
\end{itemize}

For reasons of space, all proofs appear in the Appendix.

\section{Related Works}
{\bf HJ reachability} has been widely applied to safety verification and safety-critical control of non-linear dynamical systems \cite{amodei2016concrete,hsu2023safety,borquez2024safety}. 
To mitigate the curse of dimensionality, numerous methods have been proposed to improve the scalability of HJ analysis.
{\em Decomposition techniques}, for example, exploit the structure of system dynamics to decompose high-dimensional HJ problems into multiple lower-dimensional subproblems \cite{chen2017exact,chen2018decomposition}.
{\em Adaptive discretization} and {\em efficient PDE solvers} accelerate HJ computation \cite{herbert2021scalable,bansal2017hamilton}, while neural methods approximate HJ solutions for high-dimensional systems \cite{bansal2021deepreach,yang2025scalable,feng2025bridging}.

Beyond grid-based approaches, {\em characteristics-based methods} avoid global
state-space discretization by reformulating pointwise HJ evaluation in terms of
Hamiltonian characteristics and finite-dimensional optimization
\cite{yegorov2021perspectives,chow2019algorithm,subbotina2006method}.
Our approach differs in its
computational construction: rather than solving a finite-dimensional
optimization problem independently at each queried state, we evolve a
time-varying quadratic representation of the local HJ geometry along a
reference trajectory.
More recently, local HJ computation and conservative approximation methods have
been developed to improve computational efficiency, while retaining safety
guarantees \cite{tonkens2024patching,sharpless2026conservative}.
In particular, Sharpless et al. achieve scalable nonlinear HJ reachability by constructing
conservative linear envelopes of the dynamics and applying Hopf-based computation \cite{sharpless2026conservative}.
In contrast, GEBFs retain the original nonlinear dynamics and construct a trajectory-conditioned quadratic
barrier by propagating local first- and second-order HJ geometry along a
reference trajectory.



{\bf 
Control Barrier Functions} (CBFs) have been widely used for safety-critical control \cite{wieland2007constructive,ames2019control}.
Recent research has extended the classical framework through {\em backup} \cite{chen2021backup}, {\em learning-based} \cite{cheng2019end,taylor2020learning} and {\em time-varying} barrier functions \cite{igarashi2019time,dai2025differentiable}. 
\textcolor{black}{In these approaches, any temporal variation of the barrier geometry is specified by design, rather than being generated through the underlying system dynamics.
They do not explicitly represent or propagate the evolution of the safety boundary induced by the system dynamics.}

Trajectory-based approaches have constructed local safety certificates
using polynomial or quadratic representations
\cite{tedrake2010lqr,majumdar2017funnel}.
Kumar et al.~\cite{kumar2023fast} use reach-avoid differential dynamic
programming to obtain local quadratic value approximations for constructing
implicit CBF safety filters. 
In contrast, GEBFs propagate local value, gradient, and Hessian information through HJ-derived evolution equations along reference trajectories, yielding an explicit time-varying quadratic barrier representation with established consistency properties relative to the terminal-value HJ PDE solution.
Furthermore, recent studies have investigated
theoretical connections between HJ reachability and barrier-function-based
safety certificates \cite{choi2021robust,hirsch2025viscosity}.
Building on the relationship between HJ reachability and barrier functions, our approach focuses on explicitly propagating local geometric information of the terminal-value HJ PDE solution through a time-varying quadratic barrier representation. In particular, we establish the consistency of the propagated value, gradient, and Hessian information with the HJ value function along optimal characteristics under suitable regularity assumptions.



Recent advances in AI have increasingly investigated dynamic safety through learning-based methods, including {\em safe reinforcement learning} \cite{luo2021learning,belardinelli2025probabilistic}, {\em neural barrier certificates} \cite{jin2020neural,dawson2023safe}, and {\em uncertainty-aware planning} under learned dynamics \cite{sun2023conformal}.
Rather than explicitly propagating dynamic safety boundaries, these methods typically characterize safety through learned policies, learned certificates, and uncertainty estimates.
Our work complements these methods by providing an explicit geometric representation of dynamic safety together with a theoretical characterization of its relationship to the terminal-value HJ PDE solution.

\section{Preliminaries}\label{sec_Preliminaries}

We consider autonomous agent operating in a  safety-critical dynamic environment containing moving obstacles. 

\subsection{Model and Problem 
Statement}\label{sec:problem_formulation}



\paragraph{Model Dynamics.}
Let
$\bs x_e\in\mathbb{R}^n$
and
$\bs x_o\in\mathbb{R}^n$
denote the states of the ego agent and of a dynamic obstacle, respectively.
Consider a control $u\in\mathcal U\subseteq\mathbb R^m.$
Their dynamics (Fig.~\ref{fig:illustration_environment}.a) are governed by
\begin{align}
\dot{\bs x}_e
&=
f_e(\bs x_e,u)
=
f_l(\bs x_e)
+
G(\bs x_e)u,\label{system_ego}
\\
\dot{\bs x}_o
&=
f_o(\bs x_o),\label{system_obstcale}
\end{align}
where
$f_l:\mathbb R^n\rightarrow\mathbb R^n$,
$G:\mathbb R^n\rightarrow\mathbb R^{n\times m}$, and
$f_o:\mathbb R^n\rightarrow\mathbb R^n$
are locally Lipschitz continuous. 
$f_{l},f_{o}\in C^2(\Omega;\mathbb R^n)$ and $
G\in C^2(\Omega;\mathbb R^{n\times m}).$
The control $u$ enters the agent dynamics affinely, meaning that it appears linearly in the dynamics (possibly together with a drift term) \cite{jurdjevic1997geometric, Tahmid2025ValueIteration},
and feasible control range is $\mathcal U=\prod_{i=1}^{m}[-c_i,c_i]\subseteq\mathbb R^m,
~ c_i>0$.
Throughout this paper, we assume that both the agent and dynamic obstacle evolve within a bounded statespace
$B_{\kappa}\subseteq\mathbb R^\kappa$
($0\!<\!\kappa\!\le \! n$).

For safety analysis, we reformulate the interaction between the agent and the dynamic obstacle as a coupled dynamical system.
Define the relative state $\mathbf e=\bs x_e-\bs x_o,$ relative vector field $f(\mathbf e,\bs x_o,u)
=
f_e(\mathbf e+\bs x_o,u)
-
f_o(\bs x_o).$
and introduce the augmented state $\bs z=[\mathbf e, \bs x_o]^{\top}\in \mathbb R^{2n}.$
The corresponding coupled dynamics become
\begin{align}
\dot{\bs z}
=
F(\bs z,u)
=
\begin{bmatrix}
f(\mathbf e,\bs x_o,u)\\
f_o(\bs x_o)
\end{bmatrix}.
\label{eq:coupledDynamics}
\end{align}
where $F(\cdot,u)\in C^2(\Omega;\mathbb R^n),
\forall u\in\mathcal U.
$

\textcolor{black}{For an admissible control $u(\cdot)$ and initial condition $\bs z_0$,
let $\bs z_u(\cdot)$ denote the state trajectory of
\eqref{eq:coupledDynamics} satisfying
$\bs z_u(0)=\bs z_0$.}
Safety is naturally characterized in the coupled state space.

\paragraph{Problem Statement.}
Let
$
\mathcal T=
\left\{
\bs z:
\|\mathbf e\|\le d_{\rm safe}
\right\}
$
denote the collision set,
where
$d_{\rm safe}>0$
is the prescribed safety distance.
The corresponding safe set is $\mathcal S=\mathbb R^{2n}\setminus\mathcal T.$
Given an initial state $\bs z_0\in\mathcal S$, the primary objective is to assess whether it belongs to the finite-horizon safe set
\begin{equation*}
\mathcal K_T
\!=\!\{
\bs z_0\in\mathcal S
\!\mid\!
\exists u(\cdot)\in \mathcal U
\text{ such that }
\bs z_u(t)\in\mathcal S,
\ \forall t\in[0,T]
\},
\end{equation*}
The safety-evaluation problem is therefore to determine whether $\bs z_0\in\mathcal K_T$.





This safety specification naturally defines a \emph{dynamic safety evaluation problem}:
given an initial state, determine whether the controller has a policy to maintain safety throughout the set time horizon.
As the surrounding environment evolves, the geometry of the safe set changes continuously, giving rise to a time-varying safety boundary in the coupled state space.
Motivated by this challenge, the following sections investigate a computationally efficient representation of the local geometry of the HJ value function through its characteristic evolution, without constructing a global solution over the entire state space.


\subsection{\! Control Barrier Function and HJ Reachability}

In this section, we recall the standard notion of control barrier functions and HJ reachability.
\textcolor{black}{
\paragraph{Control Barrier Functions.}
Consider the coupled dynamical system
\eqref{eq:coupledDynamics}.
Let $\mathcal C
=
\{\bs z\in\mathbb R^{2n}:h(\bs z)\ge0\},$
where
$h:\mathbb R^{2n}\rightarrow\mathbb R$
is continuously differentiable.
Following the standard control barrier function (CBF) formulation
\cite{wieland2007constructive},
$h$ is a CBF if
\begin{align}
    \bs z\in\mathcal T
    &\Longrightarrow
    h(\bs z)<0,
    \label{eq:cbf_def1}
    \\
    \forall \bs z\in\mathcal C,\;
    &\exists\,u\in\mathcal U
    \text{ such that }
    \nabla h(\bs z)^{\top}
    F(\bs z,u)
    \ge0.
    \label{eq:cbf_def2}
\end{align}}

\begin{figure}
    \centering
    \includegraphics[width=\linewidth]{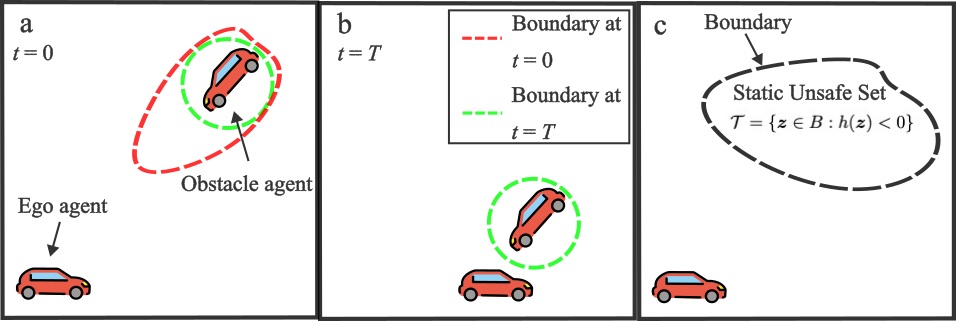}
    \vspace{-0.6cm}
    \caption{Ego agent facing dynamic obstacle agent at $t=0$ (a) and $t=T$ (b). 
    The boundary of unsafe set evolves from the red dash line at $t=0$ to the green dash line at $t=T$.
    (c) Static environment with fixed unsafe set.}
    \vspace{-0.6cm}
    \label{fig:illustration_environment}
\end{figure}
\vspace{-0.4cm}
Classical CBFs define a fixed barrier geometry in the global state space throughout the system's evolution, as illustrated by the boundary of the unsafe set in Fig.~\ref{fig:illustration_environment}.c.
In dynamic environments, however, the safety boundary evolves continuously with the underlying system dynamics.
This suggests that the barrier geometry should evolve accordingly, rather than remain fixed.
Motivated by this observation, we introduce a time-varying barrier representation whose safety boundary and local geometry co-evolve with the system dynamics (Definition~\ref{def_gebf}, Section~\ref{sec_gebf_design}).

\paragraph{HJ Reachability for Dynamic Safety Representation.} 
To characterize the evolution of safety boundary, we seek a mathematical representation of the boundary.
HJ reachability described by HJ PDE \cite{mitchell2005time,bansal2017hamilton,chen2018hamilton,ganai2024hamilton} provides a rigorous representation of dynamic safety 
through a value function (i.e., the solution of HJ reachability) defined over the state space.
The value function evolves implicitly according to the system dynamics, and 
its zero level set coincides with the boundary separating safe and unsafe states over time.

Consider an open set $\Omega\subseteq\mathbb R^{2n}.$ 
\textcolor{black}{Define the Hamiltonian} $ H(\bs z,p):=    \max_{u\in\mathcal U} p^\top F(\bs z,u)$, $(\bs z,p)\in \Omega\times\mathbb R^{2n}$. 
\textcolor{black}{Let $V:\Omega\times[0,T]\to\mathbb R$}
satisfy HJ PDE 
\begin{equation}
    \frac{\partial V}{\partial t}(\bs z,t)
    +
    H(\bs z,\nabla_{\bs z}V)
    =
    0,~(\bs z,t)\in \Omega\times(0,T), 
    \label{eq:hj_continuation}
\end{equation}
which is generally understood in the viscosity sense.

\begin{definition}[Viscosity Solution \cite{evans2022partial}]\label{def:viscosity_solution}
A continuous function
$w:\Omega\times(0,T)\rightarrow\mathbb R$
is called a \emph{viscosity supersolution} of
Eq.~\eqref{eq:hj_continuation}
if, for every test function
$\varphi\in C^\infty(\Omega\times(0,T))$,
whenever $w-\varphi$ attains a local minimum at
$(\bs z^*,t^*)$, it holds
\begin{equation*}
    \frac{\partial \varphi}{\partial t}(\bs z^*,t^*)
    +
    H\!\left(
        \bs z^*,
        \nabla_{\bs z}\varphi(\bs z^*,t^*)
    \right)
    \ge 0 .
\end{equation*}
The definition of a \emph{viscosity subsolution} is obtained analogously by replacing the local minimum with a local maximum and reversing the inequality. A \emph{viscosity solution} is a function that is both a viscosity supersolution and subsolution.
\end{definition}

Let $l:\Omega\rightarrow\mathbb R$
be the level function satisfying $\mathcal S_l
=
\{
\bs z:
l(\bs z)\ge0
\},$
where
$\mathcal S_l$
denotes the desired safe set.
For an admissible control policy
$u(\cdot)$,
define the trajectory payoff $\mathcal V(\bs z,u(\cdot))
=
\inf_{t\in[0,T]}
l\!\left(
\bs z_u(t)
\right),$
which represents the minimum safety level attained along the trajectory.
The HJ value function is defined as
\begin{equation*}\label{value_function}
V(\bs z,t)
=
\sup_{u(\cdot)\in\mathcal U}
\mathcal V(\bs z,u(\cdot)).
\end{equation*}
Accordingly, the HJ safe set is $\mathcal S_V(t)
=
\{
\bs z:
V(\bs z,t)\ge0
\},$
whose boundary is given by the zero level set $V(\bs z,t)=0.$
The HJ value function serves as the reference representation of dynamic safety throughout the remainder of the paper.

\section{CBFs as Viscosity Supersolutions}\label{sec_viscosity_sup}

The HJ value function introduced in the previous section provides a mathematical representation of the evolving safety boundary.
The following lemma establishes that, for the coupled dynamics~\eqref{eq:coupledDynamics}, safety is completely characterized by the corresponding HJ reachability without a disturbance term.
\textcolor{black}{
\begin{lemma}[Necessary and Sufficient Safety Condition]
\label{iff_safety}
Let $V$ be the viscosity solution of the
variational inequality
 \begin{align}\label{eq:HJVI}
        \!\min\left\{\!l(\bs z)\!-\!V(\bs z,t),\! \frac{\partial V}{\partial t}
\!+\!
\max_{u \in \mathcal U}
\nabla_{\bs z} V \!\cdot\! F(\bs z,u)\!
\right\}\!=\!0,
    \end{align}
with $V(\bs z,T)=l(\bs z).$
Then, for an initial state $\bs z_0$, there exists an admissible
control $u(\cdot)$ such that $l(\bs z_u(t))\ge 0,
~\forall t\in[0,T],$
if and only if $V(\bs z_0,0)\ge 0.$
\end{lemma} }

\textcolor{black}{
The variational inequality in \eqref{eq:HJVI} contains two regimes.
When $V(\bs z,t)<l(\bs z)$, the HJ equation is active and the variational inequality reduces to solving the HJ equation $$ \frac{\partial V}{\partial t}+\max_{u\in\mathcal U}
\nabla_{\bs z}V(\bs z,t)^\top F(\bs z,u)=0.$$
We refer to this regime as the \emph{HJ evolution branch}.
When $V(\bs z,t)=l(\bs z)$, the constraint is active and value function is the same as $l$; we refer to this regime as the
\emph{constraint-active branch}.
We denote by $V_{\mathrm{VI}}(\bs z,t)$ the safety-tube value of HJ variational inequality in \eqref{eq:HJVI}.
}

\begin{remark}[Scope with respect to HJ equation]
The safety-tube formulation is characterized by the HJ variational inequality in \eqref{eq:HJVI}.
In regions where $V_{\mathrm{VI}}(\bs z,t)<l(\bs z)$, the variational inequality reduces to the HJ equation. In regions where $V_{\mathrm{VI}}=l$ holds locally, the corresponding differential information is determined by the terminal function $l$, provided the required smoothness conditions hold.

In this work, we focus on the local geometric evolution governed by the terminal-value HJ PDE
\begin{equation}
    \frac{\partial V}{\partial t}
    +
    \max_{u\in\mathcal U}
    \nabla_{\bs z}V(\bs z,t)^\top F(\bs z,u)
    =0, 
    \label{eq:HJPDE}
\end{equation} 
with $V(\bs z,T)=l(z)$.
We denote its solution by $V$ to distinguish it from the safety-tube value $V_{\mathrm{VI}}$.
Our objective is to construct an explicit, trajectory-conditioned representation of the local first- and second-order HJ information without resolving the solution over the entire state space. The differential propagation results developed below concern $V$ along Hamiltonian-optimal trajectories.
\end{remark}

The relationship between control barrier functions
and HJ equation \eqref{eq:HJPDE} has been investigated.
In particular, Hirsch et al.~\cite{hirsch2025viscosity}
establish a general connection between viscosity control barrier functions and control barrier value functions under
class-$\mathcal K$ barrier conditions. 
Our subsequent construction concerns the time-dependent HJ equation~\eqref{eq:HJPDE},
without an additional class-$\mathcal K$ term. For completeness, we recall the following standard sufficient condition for a smooth function to be a viscosity supersolution of this equation.

\begin{proposition}[CBF as a Viscosity Supersolution]\label{propo_cbf_supersolution}
Let $h:\Omega\times(0,T)\rightarrow\mathbb R$
be a continuously differentiable time-varying barrier function. Suppose that
\(h\) satisfies 
\begin{equation*}
    \frac{\partial h}{\partial t}(\bs z,t)
    +
    \max_{u\in\mathcal U}
    \nabla_{\bs z}h^\top F(\bs z,u)
    \ge 0.
    \label{eq:hj_type_barrier_ineq}
\end{equation*}
Then \(h\) is a viscosity supersolution of
~\eqref{eq:hj_continuation}.
\end{proposition}

\paragraph{Local Second-Order Geometric Relation.}

Proposition~\ref{propo_cbf_supersolution} established a connection between the CBF and
HJ reachability by interpreting a smooth barrier as a viscosity supersolution.
Furthermore, this connection also can present the geometric relation between the CBF and HJ value function.

This requires some care, as HJ value functions may develop nonsmooth features, and classical Hessians are not necessarily well-defined everywhere.
We therefore use second-order
jets \cite{crandall1992user} to characterize the local second-order geometry (SOG) of HJ value function.
The formal definition and technical background are deferred to Appendix \ref{append_sec_viscosity_sup}.

The following result establishes the local second-order geometric relationship between the CBF and HJ value function.

\begin{proposition}[Local Second-order Geometric Relation]
\label{pro:local_second_order_relation}
Let $h\in C^2(\Omega\times(0,T))$ and
$V\in C(\Omega\times(0,T))$.
For $\theta>0$, define the local contact functional
$\Psi_\theta(\bs z_1,\bs z_2)
=
V(\bs z_1,t)
-
h(\bs z_2,t)
-
\frac{\theta}{2}
\|\bs z_1-\bs z_2\|^2 .$
Suppose that $\Psi_\theta$ attains a local maximum at
$(\bs z_1^\theta,\bs z_2^\theta)\in\Omega\times\Omega$ and $V(\cdot,t)$ is twice differentiable at
$\bs z_1^\theta$.
Then,
\[
\nabla_{\bs z}^{2}V(\bs z_1^\theta,t)
\preceq
\nabla_{\bs z}^{2}h(\bs z_2^\theta,t), \text{ for $t\in (0,T).$}
\]
\end{proposition}
The quantitative estimate on the separation of $\bs z^{\theta}_{1}$ and $\bs z^{\theta}_{2}$ is given in Appendix \ref{append_sec_viscosity_sup}, showing that they become arbitrarily close when the CBF locally approximates HJ value function or $\theta$ is large enough.

\begin{remark}
Proposition~\ref{pro:local_second_order_relation} provides the
theoretical motivation for relating the local SOG of a smooth barrier
function to that of the HJ value function.
While the matrix inequality characterizes a one-sided relation between
their local second-order geometries, it does not recover the local
curvature of the HJ value function itself.
Since our objective is to approximate the local HJ geometry rather than
merely preserve a one-sided curvature relation, we adopt the stronger
matching condition
\[
    \nabla_{\bs z}^{2}V
    =
    \nabla_{\bs z}^{2}h,
\]
which forms the basis of the GEBF construction developed in the
following section.
\end{remark}

\section{Geometry-Evolving Barrier Functions}\label{sec_gebf_design}

Lemma~\ref{iff_safety} shows that dynamic safety is completely characterized by the HJ value function, requiring solving a global HJ PDE \eqref{eq:HJPDE}. Instead of maintaining this global representation, our objective is to construct a compact local representation that evolves with the safety boundary.
The key idea is to propagate the local geometry of HJ value function along \emph{a reference trajectory induced by the Hamiltonian-optimal control}.
We construct a local barrier representation (Definition \ref{def_gebf}) that preserves the geometric information governing dynamic safety, as motivated by Proposition~\ref{pro:local_second_order_relation}.


\begin{definition}[GEBF]\label{def_gebf} 
A {\em Geometry-Evolving Barrier Function} is a continuously differentiable,
time-varying function $h:\mathbb{R}^{2n}\times[0,T]\rightarrow\mathbb{R},$
constructed to locally represent the evolving safety boundary around a reference trajectory.
Its associated barrier region at time $t$ is defined as $\mathcal S_h(t)
=
\left\{
\bs z\in\mathbb{R}^{2n}
:
h(\bs z,t)\ge0
\right\}.$
Throughout this paper, we consider the quadratic parametrization
\begin{align}\label{eq:gebf}
    h(\bs z,t)
=
\frac12
\bs z^\top
\bs P(t)
\bs z
+
\bs Q(t)^\top
\bs z
+
c(t),
\end{align}
where $\bs P(t)\!=\!\bs P\!^\top\!(t),
~
\bs Q(t),
~c(t)$
are time-varying functions.
\end{definition}

Throughout this paper, the proposed GEBF is evolved such that, along the reference trajectory, 
\begin{equation}
\frac{\partial }{\partial t}h(\bs z,t )
+
\max_{u\in\mathcal{U}}\nabla_{\bs z}h^\top(\bs z,t ) F(\bs z,u)
\ge0.
\label{eq:gebf_condition}
\end{equation}

\begin{remark}
Having introduced the GEBF, the key remaining task is to determine how parameters
$\bs P(t)$,
$\bs Q(t)$,
and
$c(t)$
should evolve.
Guided by Proposition~\ref{pro:local_second_order_relation},
we construct the GEBF by progressively matching the local geometric information of HJ value function.
Specifically,
function
$\bs P(t)$
is first evolved to capture the local SOG,
followed by the construction of
$\bs Q(t)$
and
$c(t)$
to match the corresponding first-order and zeroth-order information.
\end{remark}

Since HJ value functions may exhibit nonsmooth behavior, the differential relations considered below are formulated in the almost-everywhere sense along the relevant trajectories, under the following regularity assumptions.


\begin{definition}[Semiconvexity \cite{ambrosio1993propagation}]
Let $\Omega\subseteq\mathbb R^{2n}$ be an open convex set. A function
$\kappa:\Omega\to\mathbb R$ is {\em semiconvex} on $\Omega$ if there exists a constant
$C\ge0$ such that $\kappa(\bs z)+\frac{C}{2}\|\bs z\|^2$
is convex on $\Omega$.
\end{definition}

\begin{assumption}
\label{ass:semiconvex_valuefunc}
For each $t\in[0,T]$, the value function $V(\cdot,t)$ is semiconvex on
$\Omega$ and locally Lipschitz continuous.
\end{assumption}

\begin{assumption}
\label{ass:hamilton_maximizer}
The Hamiltonian maximizer $$u^*(\bs z,t)
\in
\arg\max_{u\in\mathcal U}
\nabla_{\bs z}V(\bs z,t)^\top F(\bs z,u)$$
is unique almost everywhere in $\Omega\times[0,T]$.
Moreover, the switching surfaces of $u^*$ are smooth hypersurfaces with
non-vanishing gradients.
\end{assumption}

\begin{remark}
These assumptions are mild regularity assumptions that are widely used in the analysis of optimal control \cite{evans1989hamiltonian} and HJ PDE \cite{evans2022partial}.
Assumption~\ref{ass:semiconvex_valuefunc} implies twice differentiability almost everywhere by Alexandrov's theorem. 
Moreover, Assumption~\ref{ass:hamilton_maximizer} is satisfied by many benchmark control-affine systems commonly used in robotics and autonomous driving, including single-integrator, Dubins-car, and unicycle models \cite{ames2016control, yu2023sequential,chen2025int2planner}.
This assumption shows that switching occurs only on a zero-measure set (Lemma 3 in Appendix \ref{append_sec_main}).
\end{remark}

\begin{assumption}\label{ass:hj_second_order}
Let $\bs z_{u}^*(t)$ be the HJ-optimal
characteristic under consideration.
Away from the switching times, assume that the HJ value function $V(z,\cdot)\in C^{1}$ and 
$V(\cdot,t)\in C^{2}$ in a neighborhood of the optimal characteristic.
Furthermore, assume that $\nabla_{\bs z} V(\bs z_{u}^*(t),t),$
$\nabla_{\bs z}^2 V(\bs z_{u}^*(t),t),$
admit the absolutely continuous extension
to $[0,T]$.
\end{assumption}

We use \textit{absolute continuity} (AC) in its standard sense \cite{royden1988real}:
$m\in AC([0,T])$ admits an almost-everywhere derivative
$m'\in L^1([0,T])$ and satisfies
$m(t)=m(0)+\int_0^t m'(s)\,ds$
.

\begin{remark}
The local second-order regularity in Assumption \ref{ass:hj_second_order}
is consistent with established sensitivity-based
regularity results for HJ value
functions, e.g.~\cite{cannarsa2015second}.
In the present control-affine setting, the
maximized Hamiltonian is generally nonsmooth
across switching surfaces. However, away from
these surfaces, it coincides locally with a
smooth Hamiltonian branch.
Although the optimal control may exhibit switching between different branches, such switching does not necessarily imply a loss of regularity of the HJ value function along the optimal characteristic. We therefore focus on trajectories along which the gradient and Hessian admit absolutely continuous extensions over the entire time horizon. The numerical results presented below illustrate behavior consistent with this regularity assumption.
\end{remark}

\subsection{Evolution of the Second-Order Geometry}
We first derive the evolution of the local Hessian (curvature) of the HJ value function along a Hamiltonian-optimal trajectory.
We refer to this trajectory as the HJ-optimal trajectory. 


\begin{theorem}[First- and Second-Order Dynamics of HJ Value Function]
\label{thm:hessian_dynamics_value}
Consider HJ PDE in Eq.~\eqref{eq:HJPDE}.
Let $$u^*(t)
\in
\arg\max_{u\in\mathcal U}
\nabla_{\bs z}V(\bs z,t)^\top F(\bs z,u),$$
and let $\bs z_{u}^*(t)$ be a reference trajectory (i.e., HJ-optimal trajectory) satisfying $\dot{\bs z}_{u}^*(t)
=
F(\bs z_{u}^*(t),u^*(t)).$
At points where $V$ is twice differentiable, define $\bs H(t)
=
\nabla_{\bs z}^2 V(\bs z_{u}^*(t),t)$ and $\bs g(t)=\nabla_{\bs z} V(\bs z_{u}^*(t),t)$.
Assumption \ref{ass:semiconvex_valuefunc}, \ref{ass:hamilton_maximizer} and \ref{ass:hj_second_order} are satisfied.
Then, for almost every $t\in[0,T]$, the Jacobian satisfies
\begin{equation}
\dot{\bs g}(t)
=
-\bs J(t)^\top\bs g(t),
\label{eq:value_jacobian_evolution}
\end{equation}
and the Hessian satisfies
\begin{equation*}
\dot{\bs H}(t)
=
-\bs H(t)\bs J(t)
-
\bs J(t)^\top\bs H(t)
-
\bs \Delta(t),
\label{eq:hessian_riccati_value}
\end{equation*}
where $
\bs J(t)
=
\nabla_{\bs z}F(\bs z_{u}^*(t),u^*),$ and 
$\bs \Delta(t)
=
\nabla_{\bs z}
\!\left[
\bs J(t)^\top
\right]
\nabla_{\bs z}V(\bs z_{u}^*(t),t)$
collects the second-order variation of the vector field.
\end{theorem}

Theorem~\ref{thm:hessian_dynamics_value} provides the evolution law of the
local SOG of HJ value function.
By Proposition~\ref{pro:local_second_order_relation}, we enforce
\[
\nabla_{\bs z}^{2}V(\bs z_u^*,t)
=
\nabla_{\bs z}^{2}h(\bs z_u^*,t),
\]
along trajectory $\bs z_u^*$.
Therefore, $\bs P(t)$ in Eq.~\eqref{eq:gebf} evolves according to Eq.~\eqref{eq:hessian_riccati_value}.
We define a Hessian evolution equation
\begin{equation}
\!\dot{\bs P}(t)
=
-\bs P(t)\bs J(t)
-
\bs J(t)^\top\bs P(t)
-
\bs \Delta_P(t),~t\in [0,T]
\label{eq:P_evolution}
\end{equation}
where $\bs \Delta_P(t)=\nabla_{\bs z}
\!\left[
\bs J(t)^\top
\right]
\nabla_{\bs z}h(\bs z^{*}_{u}(t),t)$.
Consider the level function $l(\bs z)$ as the terminal condition $h(\bs z, T)=l(\bs z).$

Based on the backward propagation of~\eqref{eq:P_evolution}, we can ensure the matrix-valued function $\bs P(t)$ is always the same as the 
Hessian matrix of HJ value function along trajectory $\bs z^*_{u}$ if the value function is smooth along trajectory.

\begin{theorem}[Local Second-Order Geometric Consistency]
\label{thm:second_order_consistency}
Let $\bs z_{u}^*(t)$ be a continuously differentiable reference trajectory on
$[0,T]$.
Let $\bs P(t)$ satisfy Eq.~\eqref{eq:P_evolution} with terminal condition $\bs P(T)=\bs H(T).$
Assumption \ref{ass:semiconvex_valuefunc}, \ref{ass:hamilton_maximizer} and \ref{ass:hj_second_order} are satisfied.
If $\bs\Delta_P(t)=\bs\Delta(t),$ for a.e. $t\in[0,T],$
then 
$\bs P(t)=\bs H(t),
~
\text{for a.e. } t\in[0,T].$
\end{theorem}

\begin{remark}
Theorem~\ref{thm:second_order_consistency} establishes that the proposed
evolution law for $\bs P(t)$ is consistent with the local Hessian dynamics of
HJ value function.
Therefore, once 
$\bs\Delta(t)=\bs\Delta_{P}(t)$ is guaranteed through the Jacobian matching procedure introduced next, the resulting matrix $\bs P(t)$ exactly reproduces the local SOG of HJ value function along the reference trajectory.
Consequently, $\bs P(t)$ serves as the second-order geometric component of the
proposed GEBF.
\end{remark}

The above consistency is a property along the HJ-optimal
trajectory $\bs z_{u}^*$. Since $V$ is only assumed to be semiconvex
and locally Lipschitz, the equality is understood almost everywhere
along $\bs z_{u}^*$ where the corresponding second-order derivative exists.
No classical Hessian representation is implied at points where $V$
is not twice differentiable.

\subsection{Evolution of the First-Order Geometry}
The previous subsection matched the SOG of the GEBF to that of the HJ value function. However, SOG alone is insufficient to recover the local control behavior of HJ reachability.


Since the Hamiltonian depends explicitly on the gradient of HJ value function, matching the Hessian alone is insufficient to recover the optimal control.
The gradient of the GEBF should therefore also evolve consistently with that of the HJ value function.


Along trajectory $\bs z_u^*(t),$
the first-order geometry (FOG) satisfies Eq.~\eqref{eq:value_jacobian_evolution}.
For the GEBF~\eqref{eq:gebf},
its Jacobian is
\begin{equation}\label{eq:jacobian}
    \nabla_{\bs z}h(\bs z,t)
=
\bs P(t)\bs z+\bs Q(t).
\end{equation}
Since $\bs P(t)$ has already been determined, the remaining first-order degree of freedom is $\bs Q(t)$.

We therefore choose $\bs Q(t)$ so that  along
trajectory $\bs z_u^*$, the GEBF Jacobian $\bs h_J(t)
=
\nabla_{\bs z}h(\bs z_{u}^*(t),t)
=
\bs P(t)\bs z_{u}^*(t)+\bs Q(t),$
satisfies the same Jacobian evolution \eqref{eq:value_jacobian_evolution}:
\begin{equation}
\dot{\bs h}_J(t)
=
-\bs J(t)^\top \bs h_J(t),
\qquad
t\in[0,T].
\label{eq:gebf_jacobian_evolution}
\end{equation}

\begin{theorem}[Local FOG Consistency]
\label{thm:first_order_consistency}
Let $\bs z_{u}^*(t)$ be a continuously differentiable reference trajectory on
$[0,T]$ generated by the Hamiltonian-maximizing control $u^*(t)$.
Let $\bs h_J(t)=\nabla_{\bs z}h(\bs z_{u}^*(t),t)$ satisfy
Eq.~\eqref{eq:gebf_jacobian_evolution} with the same terminal condition
$\bs h_J(T)
=
\nabla l(\bs z^*(T)).$ Assumption \ref{ass:semiconvex_valuefunc}, \ref{ass:hamilton_maximizer} and \ref{ass:hj_second_order} are satisfied.
Then 
$$\bs h_{J}(t)=\bs g(t)~
\text{for a.e. }t\in[0,T].$$

\end{theorem}

\begin{remark}
Theorem~\ref{thm:first_order_consistency} shows that the GEBF Jacobian matches that of the HJ value function along trajectory $\bs z_{u}^*(t)$.
Accordingly, $\bs Q(t)$ is evolved to preserve the FOG and the corresponding Hamiltonian-maximizing control direction.
\end{remark}


\subsection{Evolution of the Zeroth-Order Geometry}
The previous two subsections determine the parameters
$\bs P(t)$
and
$\bs Q(t)$,
which match the SOG and FOG of the HJ value function, respectively.
The remaining scalar component
$c(t)$
is then constructed to match the value evolution of the HJ value function along trajectory  $\bs z_{u}^*(t)$.

Since Eq~\eqref{eq:jacobian}
and $\frac{\partial h}{\partial t}(\bs z,t)
=
\frac12\bs z^\top\dot{\bs P}(t)\bs z
+
\dot{\bs Q}(t)^\top\bs z
+
\dot c(t),$
we choose $c(t)$ so that $h$ satisfies the same value evolution equation as
the HJ value function along $\bs z_u^*(t)$:
\begin{equation}
\frac{\partial h}{\partial t}(\bs z_u^*(t),t)
+
\nabla_{\bs z}h(\bs z_u^*(t),t)^\top
F(\bs z_u^*(t),u^*(t))
=
0
\label{eq:h_value_evolution}
\end{equation}
Substituting Eq.~\eqref{eq:gebf} into
Eq.~\eqref{eq:h_value_evolution} gives the evolution law
\begin{align}
\dot c(t)
=&
-\frac12
{\bs z_u^*(t)}^\top
\dot{\bs P}(t)
\bs z_u^*(t)
-
\dot{\bs Q}(t)^\top
\bs z_u^*(t)
\nonumber\\
&-
\big(
\bs P(t)\bs z_u^*(t)+\bs Q(t)
\big)^\top
F(\bs z_u^*(t),u^*(t))
\label{eq:c_evolution}
\end{align}
The terminal condition $h(\bs z_u^*(T),T)
=
l(\bs z_u^*(T))$
yields
\begin{equation}
\!\!\!c(T)
\!=\!
l(\!\bs z_u^*(T)\!)
-
\frac12
{\bs z_u^*(T)}^\top\!
\bs P(T)
\bs z_u^*(T)
\!-
\bs Q(T\!)^\top\!
\bs z_u^*(T)
\label{eq:c_terminal_condition}
\end{equation}

\begin{theorem}[Zeroth-Order Geometric Consistency]
\label{thm:zeroth_order_consistency}
Let $\bs z_u^*(t)$ be a continuously differentiable reference trajectory on
$[0,T]$ generated by the Hamiltonian-maximizing control $u^*(t)$.
Suppose that the FOG  consistency holds along
$\bs z_u^*(t)$, namely $\nabla_{\bs z}h(\bs z_u^*(t),t)
=
\nabla_{\bs z}V(\bs z_u^*(t),t),$ for a.e. $t\in[0,T].$
Assume that $c(t)$ satisfies Eq.~\eqref{eq:c_evolution} with the terminal
condition~\eqref{eq:c_terminal_condition}.
Assumption \ref{ass:semiconvex_valuefunc}, \ref{ass:hamilton_maximizer} and \ref{ass:hj_second_order} are satisfied.
Then, 
$$h(\bs z_u^*(t),t)
=
V(\bs z_u^*(t),t),
~
\text{for a.e. }t\in[0,T].$$
\end{theorem}

\begin{remark}
Together with Theorems~\ref{thm:second_order_consistency}
and~\ref{thm:first_order_consistency},
Theorem~\ref{thm:zeroth_order_consistency} shows that the GEBF
matches HJ value function along trajectory $\bs z_u^*(t)$ at the
second-, first-, and zeroth-order levels.
\end{remark}

\section{Algorithmic Construction}\label{sec:algorithm}

The developments above assume that the reference trajectory is induced by the Hamiltonian-optimal control of HJ value function.
In practice, however, such a trajectory is unavailable a priori because the Hamiltonian-optimal controller itself depends on the evolving GEBF (Algorithm~\ref{alg:GEBF}).
We therefore initialize the procedure with a nominal trajectory generated by an arbitrary admissible control and employ an alternating iterative scheme that jointly refines the GEBF and the reference trajectory


Algorithm~\ref{alg:GEBF} summarizes the proposed procedure. Starting from an initial nominal trajectory (Line 1),
each iteration computes the Hamiltonian-maximizing control and the corresponding local Jacobian (Line~6--8), and then successively updates the parameter $\bs P(t)$, $\bs Q(t)$, and $c(t)$,
according to Theorems~\ref{thm:second_order_consistency},
\ref{thm:first_order_consistency},
and
\ref{thm:zeroth_order_consistency}, respectively (Line~10).

The resulting GEBF generates an updated Hamiltonian-optimal trajectory (Line 12--16), which becomes the reference trajectory for next iteration. 
Both the local safety geometry and reference trajectory are progressively refined until convergence or the maximum number of iterations is reached.

It is important to distinguish the HJ-optimal trajectory used in
the preceding consistency analysis from the reference trajectories
generated during the iteration. 
In general, $\bs z_{\mathrm{ref}}^k$ need not be an HJ-optimal trajectory.
Accordingly, the propagated matrices $\bs P^{k+1}$ and
$\bs Q^{k+1}$ at an intermediate iteration describe the local geometry constructed relative to the current reference trajectory;
they are not asserted to coincide with the Hessian and gradient of
the HJ value function along $\bs z_{\mathrm{ref}}^k$.
Nevertheless, Theorem~\ref{thm:trajectory_lower_bound} establishes a trajectory-wise one-sided relation between the GEBF and HJ value function under its stated conditions, without requiring the reference trajectory to be HJ-optimal.

\begin{remark}
The proposed algorithm alternates between GEBF construction and reference trajectory refinement.
Unlike HJ reachability, which solves a global HJ PDE, 
the proposed framework propagates only local geometric information along a reference trajectory.
Its computational cost therefore depends primarily on the trajectory length rather than state-space discretization, improving scalability for high-dimensional systems.
This efficiency is achieved by replacing the global HJ representation with a local geometric representation, which captures only the local safety geometry around the reference trajectory.
For a different operating region, the framework can be reinitialized from
another nominal trajectory to construct a new locally consistent GEBF.
   
\end{remark}

\begin{remark} 
\textcolor{black}{
Although no formal convergence guarantee for Algorithm~1 is established, the local consistency results in Section~\ref{sec_gebf_design} provide insight into the observed iterative behavior.
Once the reference trajectory lies sufficiently close to a HJ-optimal trajectory, the backward propagation Eqs.~\eqref{eq:P_evolution}, \eqref{eq:gebf_jacobian_evolution}, and \eqref{eq:c_evolution} are solved along nearby trajectories. Since these evolution equations vary continuously with the reference trajectory, 
the propagated quadratic coefficients are expected to remain close to those of the HJ-optimal trajectory, allowing the constructed GEBF to continue providing an accurate local approximation of the HJ value, gradient, and Hessian.
The induced Hamiltonian-maximizing controls remain close to the locally optimal ones, leading to progressively smaller trajectory updates. 
This behavior is consistent with the numerical experiments,
where the trajectory-update error rapidly decreases and stabilizes after
approximately 20 iterations  (Fig. \ref{fig:error_iteration}, Appendix \ref{append_ex:8d_simple_car_noRela}).
}
\end{remark}

{
\setlength{\textfloatsep}{4pt}
\setlength{\intextsep}{4pt}
\begin{algorithm}[t!]
\caption{Iterative GEBF Construction}
\label{alg:GEBF}

\textbf{Input}: Initial state $\bs z_0$, horizon $[0,T]$,
terminal function $l(\bs z)$, time grid
$0=t_0<t_1<\cdots<t_N=T$, maximum iteration number $K$,
and tolerance $\epsilon$.

\textbf{Output}: GEBF $h(\bs z,t)$ 
and the refined trajectory $\bs z_{\mathrm{ref}}(t)$.

\begin{algorithmic}[1]

\STATE Generate the initial reference trajectory
$\bs z_{\mathrm{ref}}^0(t)$ \textcolor{black}{generated by any feasible controller $u_{\mathrm{nom}}^0(t)$ (e.g., a constant, heuristic, PID, or MPC controller)}

\FOR{$k=0,1,\ldots,K-1$}

    \STATE Set $ \bs z_i^k
    =
    \bs z_{\mathrm{ref}}^k(t_i),
    ~i=0,\ldots,N.$
    
    \STATE Initialize the terminal GEBF geometry $ \bs P_N^{k+1}$, $\bs Q_N^{k+1},$ and $ c_N^{k+1}$ based on $l(\bs z_N^k).$
    

    \FOR{$i=N,N-1,\ldots,1$}

        \STATE Evaluate the GEBF gradient at the current
        reference state: $ \bs g_i^{k+1}
        =
        \bs P_i^{k+1}\bs z_i^k
        +
        \bs Q_i^{k+1}.$
    
        \STATE Compute the Hamiltonian-maximizing control $u_i^{\!*,k\!}$.

        \STATE Evaluate the local Jacobian: $\bs J_i^k
        =
        \frac{\partial F}{\partial \bs z}
        \bigl(\bs z_i^k,u_i^{*,k}\bigr).$

        \STATE Evaluate $\bs\Delta_i^k
        \!\!=\!
        \sum_{j=1}^{n}\!
        [\bs g_i^{k+1}]_j
        \!\nabla_{\bs z}^2
        F_j(\bs z_i^k,u_i^{*,k})$. 

        \STATE Construct $\bs P_{i-1}^{k+1}$, $\bs Q_{i-1}^{k+1}$, and $c_{i-1}^{k+1}$ by solving equation \eqref{eq:P_evolution},
        \eqref{eq:gebf_jacobian_evolution}, and
        \eqref{eq:c_evolution} backward,
        


    \ENDFOR

    \STATE Construct the updated GEBF: $ h^{k+1}(\bs z,t)
    =
    \frac{1}{2}\bs z^\top
    \bs P^{k+1}(t)\bs z
    +
    {\bs Q^{k+1}(t)}^\top\bs z
    +
    c^{k+1}(t).$

    \STATE Generate the refined trajectory $\bs z_{\mathrm{ref}}^{k+1}(t)$
by forward simulation under the controller induced by $h^{k+1}$
    

    \IF{$
    \displaystyle
    \max_{t\in[0,T]}
    \left\|
    \bs z_{\mathrm{ref}}^{k+1}(t)
    -
    \bs z_{\mathrm{ref}}^k(t)
    \right\|
    <\epsilon
    $}

        \STATE \textbf{break}.

    \ENDIF

\ENDFOR

\STATE \textbf{return}
$h(\bs z,t)=h^{k+1}(\bs z,t)$ and
$\bs z_{\mathrm{ref}}(t)=
\bs z_{\mathrm{ref}}^{k+1}(t)$.
\vspace{-0.1cm}
\end{algorithmic}
\end{algorithm}
}\vspace{-1ex}

A fundamental consistency question is whether the HJ-optimal trajectory is preserved by this update.
In particular, if the reference trajectory coincides with an HJ-optimal trajectory , the GEBF construction should
induce the same Hamiltonian-maximizing control and hence reproduce the same trajectory after the forward update.
The following result establishes this
fixed-point consistency of Algorithm~\ref{alg:GEBF}. Importantly, the result
is not about the convergence of the iteration from an arbitrary initial
trajectory; rather, it identifies the HJ-optimal trajectory as a fixed
point of the trajectory-update operator.

\begin{theorem}[
Fixed point of GEBF iteration]
\label{thm:fixed_point_consistency}
Suppose Assumption \ref{ass:semiconvex_valuefunc}, \ref{ass:hamilton_maximizer} and \ref{ass:hj_second_order} hold.
Let $V$ satisfy the HJ PDE \eqref{eq:HJPDE}
with terminal condition $V(\bs z,T)=l(\bs z).$
Let $\bs z_{u}^*(t)$ be an HJ-optimal trajectory satisfying
\begin{equation*}
    \dot{\bs z}_{u}^*(t)
    =
    \bs F(\bs z_{u}^*(t),u^*(t)),
    \qquad
    \bs z_{u}^*(0)=\bs z_0,
\end{equation*}
where $u^*(t)
    \in
    \arg\max_{u\in\mathcal U}
    \nabla_{\bs z}V(\bs z^*(t),t)^\top
    \bs F(\bs z^*(t),u).$
Then $\bs z_{u}^*(t)$ is a fixed point of the trajectory-update operator
$\mathcal T$ defined by Algorithm~\ref{alg:GEBF}, i.e.,
\begin{equation*}
    \mathcal T(\bs z_{u}^*)=\bs z_{u}^*.
\end{equation*}
Equivalently, if the reference trajectory supplied to
Algorithm~\ref{alg:GEBF} is $\bs z_{u}^*(t)$, the trajectory generated by
the resulting GEBF-induced control coincides with $\bs z^*(t)$.
\end{theorem}


\textcolor{black}{\begin{remark}
Theorem~\ref{thm:fixed_point_consistency} establishes a fixed-point consistency of
the HJ-optimal trajectory with the GEBF trajectory update. 
In particular, it shows that once the reference trajectory coincides
with $\bs z^*$, the GEBF-induced update preserves it.
\end{remark}}

By construction of the scalar term $c(t)$, the GEBF satisfies
the HJ PDE along its reference trajectory $\bs z_{\mathrm{ref}}$, i.e.,
\begin{equation}\label{residual_zero}
    \frac{\partial h}{\partial t}(\bs z_{\mathrm{ref}}(t),t)
    +
    H\!\left(
        \bs z_{\mathrm{ref}}(t),
        \nabla_{\bs z}h(\bs z_{\mathrm{ref}}(t),t)
    \right)
    =0
\end{equation}
for almost every $t\in[0,T]$ at which the involved quantities
are differentiable.
The GEBF calculated by Algorithm~\ref{alg:GEBF} satisfies Eq. \eqref{residual_zero}, while it has the following property.
\vspace{-0.2cm}


\begin{theorem}[Lower bound along a reference trajectory]
\label{thm:trajectory_lower_bound}
Suppose the following conditions hold:
\begin{enumerate}
    \item Let $\bs z_{\mathrm{ref}}(t)$ be a GEBF-induced reference trajectory satisfying
    \begin{equation*}
        \dot{\bs z}_{\mathrm{ref}}(t)
        =
        \bs F(\bs z_{\mathrm{ref}}(t),
              u_{\mathrm{ref}}(t)),
    \end{equation*}
    where $u_{\mathrm{ref}}(t)\in
        \arg\max_{u\in\mathcal U}
        \nabla_{\bs z}h(
            \bs z_{\mathrm{ref}}(t),t
        )^\top
        \bs F(\bs z_{\mathrm{ref}}(t),u)$
    for almost every $t\in[0,T]$.
    \item The HJ value function $V$ satisfies
    \eqref{eq:HJPDE} with terminal condition
    $V(\bs z,T)=l(\bs z)$.
    Along $\bs z_{\mathrm{ref}}(t)$, $V$ is
    differentiable in space and time almost
    everywhere, satisfies the HJ PDE at these
    points, and
    $t\mapsto V(\bs z_{\mathrm{ref}}(t),t)$
    is absolutely continuous.
    \item The GEBF construction satisfies 
\begin{equation*}
    \frac{\partial h}{\partial t}
    (\bs z_{\mathrm{ref}}(t),t)
    +
    H\!\left(
        \bs z_{\mathrm{ref}}(t),
        \nabla_{\bs z}h(
            \bs z_{\mathrm{ref}}(t),t)
    \right)
    =0
\end{equation*}
for almost every $t\in[0,T]$, together with
the terminal matching condition $h(\bs z_{\mathrm{ref}}(T),T)
    =
    V(\bs z_{\mathrm{ref}}(T),T)
    =
    l(\bs z_{\mathrm{ref}}(T)).$
\end{enumerate}
Then, for every $t\in[0,T]$,
\begin{equation*}
    V(\bs z_{\mathrm{ref}}(t),t)
    -
    h(\bs z_{\mathrm{ref}}(t),t)
    =
    \int_t^T \delta_H(s)\,ds,
\end{equation*}
where
\begin{equation*}
\begin{aligned}
    \delta_H(t)
    &:=
    H\!\left(
        \bs z_{\mathrm{ref}}(t),
        \nabla_{\bs z}V(\bs z_{\mathrm{ref}}(t),t)
    \right)
    \\
    &\quad -
    \nabla_{\bs z}V(\bs z_{\mathrm{ref}}(t),t)^\top
    \bs F(\bs z_{\mathrm{ref}}(t),u_{\mathrm{ref}}(t))
    \ge 0 , \text{ a.e } t\in [0,T].
\end{aligned}
\end{equation*}
In particular, we have 
\begin{equation*}
    h(\bs z_{\mathrm{ref}}(t),t)
    \le
    V(\bs z_{\mathrm{ref}}(t),t),
    \qquad
    t\in[0,T].
\end{equation*}
\end{theorem}

\begin{remark}
Theorem~\ref{thm:trajectory_lower_bound} provides a trajectory-wise soundness
guarantee for the GEBF representation. In particular, along a
GEBF-induced reference trajectory satisfying the conditions of the
theorem,
\[
    h(\bs z_{\mathrm{ref}}(t),t)\ge 0
    \quad\Longrightarrow\quad
    V(\bs z_{\mathrm{ref}}(t),t)\ge 0.
\]
Hence, a state conservatively evaluated by the GEBF with respect to the terminal-value HJ solution.
This guarantee is one-sided and trajectory-wise; it does not imply a global ordering between $h$ and $V$ over the entire state space.
\end{remark}

Together, these results establish a direct connection between the GEBF construction and the HJ value function: the fixed-point property demonstrates the consistency of the geometry–trajectory update with HJ-optimal trajectories, while the one-sided relation characterizes the discrepancy between the GEBF and HJ value function along GEBF-induced reference trajectories.

\section{Empirical Evaluation}\label{sec:experiments}


This section evaluates GEBFs on representative dynamic collision-avoidance problems, using the HJ value functions and 
their derivatives numerically computed with helperOC
\cite{mitchell2005toolbox,fisac2015reach} as references. 
\textcolor{black}{
Specifically, the reference solutions are computed from the terminal-value HJ PDE \eqref{eq:HJPDE}, consistent with the differential propagation theory developed in this work.
rather than from the complete safety-tube variational inequality.
} 
We assess (i) the accuracy of GEBFs in approximating the local HJ value function; (ii) their ability to generate safe trajectories through iterative geometry evolution; and (iii) the computational cost of GEBF construction relative to grid-based HJ PDE computation.

All numerical experiments are executed on a MacBook Pro equipped with an Apple M4 chip and 24 GB of Unified Memory, running MATLAB R2026a.

\subsection{Example: $2$D Coupled System}\label{append_experiments_example:one}

\paragraph{Environment.} 
We consider a $2$D agent--obstacle collision-avoidance system.
The ego agent is governed by a single-integrator \cite{ames2016control}, while the obstacle follows a nonlinear autonomous dynamics \cite{strogatz2024nonlinear}
\begin{align}
\dot{x}_e = u,~
\dot{x}_o = -v_o+\beta\tanh(x_o),
\label{eq:ego_obs_dynamics}
\end{align}
where $x_e,x_o\in\mathbb R$ denote the positions of the agent and the obstacle, respectively,
$u\in\mathcal U=[-1,1]$ is the control input,
$v_o=0.6$ is the nominal obstacle velocity, and
$\beta=0.8$ characterizes the nonlinear drift.
The agent and obstacle start at the initial position $x_{e}=-2$ and $x_o=-1.3$, respectively.
Following Sec.~\ref{sec:problem_formulation}, we introduce the relative state $e=x_e-x_o$, and define the coupled state $\bs z=[e,x_o]^\top\in\mathbb R^2.$
The corresponding terminal condition is chosen as $l(\bs z)
=
e^2-R^2,$
where $R=0.3$ is the prescribed safety radius.
We construct the GEBF in the form~\eqref{eq:gebf}.

The iterative procedure is initialized using a nominal trajectory generated by a simple heuristic control policy.
During the initialization, the initial controller steers the agent according to the relative configuration of both the agent and the obstacle, while the trajectory is constrained to remain inside the computational domain and is terminated once the collision boundary is reached.
\begin{figure}[t!]
    \centering
    \includegraphics[width=\linewidth]{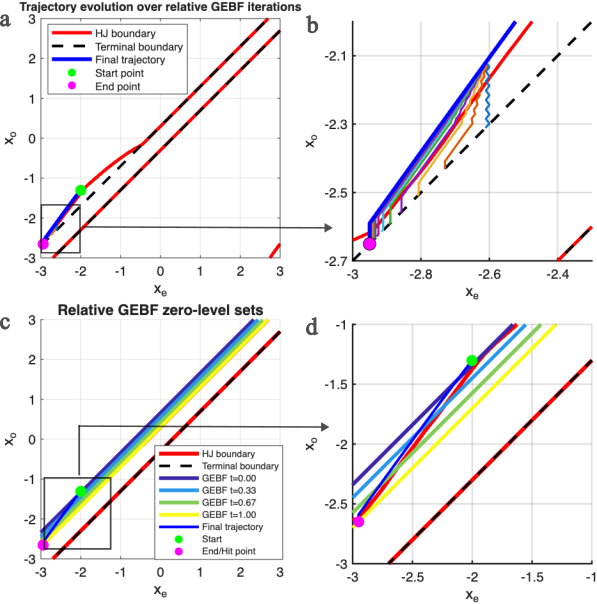}
    \vspace{-0.4cm}
    \caption{(a) Trajectory evolution during the iterative procedure. (b) Enlarged view showing the trajectory evolution. (c) Evolution of the GEBF zero-level sets. (d) Enlarged view illustrating the approximation of the zero-level set between the GEBF and HJ value function (red line) at $t=0$. The black dash line is the zero-level set of $l(\bs z).$
    The value of $V$ at start point (green dot) is $0.0523,$ which is close to the zero level set.}
    \vspace{-0.4cm}
    \label{fig:case1_all}
\end{figure}

\begin{figure}[htb!]
    \centering    \includegraphics[width=\linewidth]{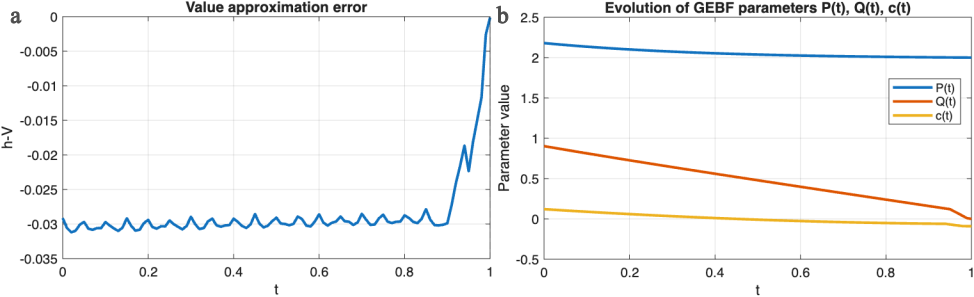}
    \vspace{-0.4cm}
    \caption{(a) Difference between the GEBF value and the HJ value function at the starting point.
    (b) The Evolution of GEBF parameters.}
     \vspace{-0.6cm}
    \label{fig:case1_rest}
\end{figure}

\paragraph{Results.} 

Fig.~\ref{fig:case1_all} demonstrates the coupled iterative refinement between the trajectory and the GEBF. Beginning with an initial trajectory, the proposed algorithm alternates between updating the local GEBF geometry and recomputing the trajectory. As shown in Figs.~\ref{fig:case1_all}(a)--(b), the iterated trajectories become progressively closer to the final trajectory, with the largest update occurring in the first iteration and only minor adjustments thereafter. 

Simultaneously, Figs.~\ref{fig:case1_all}(c)--(d) illustrate the evolution of the GEBF zero-level sets along the final updated trajectory. As time progresses, the zero-level sets evolve continuously and remain in close agreement with the zero-level set of the HJ value function at $t=0,$ which also can be validated by Fig.~\ref{fig:case1_rest}(a). 
The one-sided relationship can be observed in Fig.~\ref{fig:case1_rest}(a).
The parameters' evolutions are shown in Fig.~\ref{fig:case1_rest}(b).
Moreover, the gradient and Hessian vary smoothly along the trajectory, providing numerical evidence consistent with Assumption 3 (Fig.\ref{fig:case1_1D_gradient_ac}-\ref{fig:case1_1D_Hessian_ac}, Appendix).

\subsection{3D Relative Kinematic Car}
 \label{example:three_dimension}
\paragraph{Environment.}
We consider a planar collision-avoidance problem between a controlled ego vehicle and a non-adversarial moving obstacle.
Let $p^e, p^o\in \mathbb{R}^2$ denote the positions, and $\theta_e, \theta_o$ the heading angles, of the ego and obstacle agents, respectively.
The ego vehicle follows the constant-speed 3D kinematic car model (Appendix~\ref{append_experiments_example:three}) 
, a standard benchmark in helperOC~\cite{mitchell2005toolbox,fisac2015reach} and autonomous driving~\cite{yu2023sequential,chen2025int2planner}, while the obstacle follows the same dynamics 
with constant speed and fixed heading.
The collision set is $\mathcal L=
\left\{
(p^e,p^o,\theta_e,\theta_o):
\|p^o-p^e\|\le R
\right\},$
where $R$ denotes the safety distance.
To enable direct comparison with helperOC, which is limited to systems of at most five dimensions, we transform the original 6D system into the 5D relative-coordinate model \eqref{eq:air3d_nodisturbance} in Appendix~\ref{append_experiments_example:three}.
Here $(x_1,x_2)$ denote the obstacle position in the ego body frame and $\theta$ is the relative heading.
The terminal function becomes
$l(x_1,x_2,\theta)=x_1^2+x_2^2-R^2$.
We construct the GEBF in the form of Eq.~\eqref{eq:gebf}, and the parameter settings are provided in Appendix~\ref{append_experiments_example:three}.

\vspace{-0.1cm}
\begin{figure}[htb!]
    \centering
    \includegraphics[width=\linewidth]{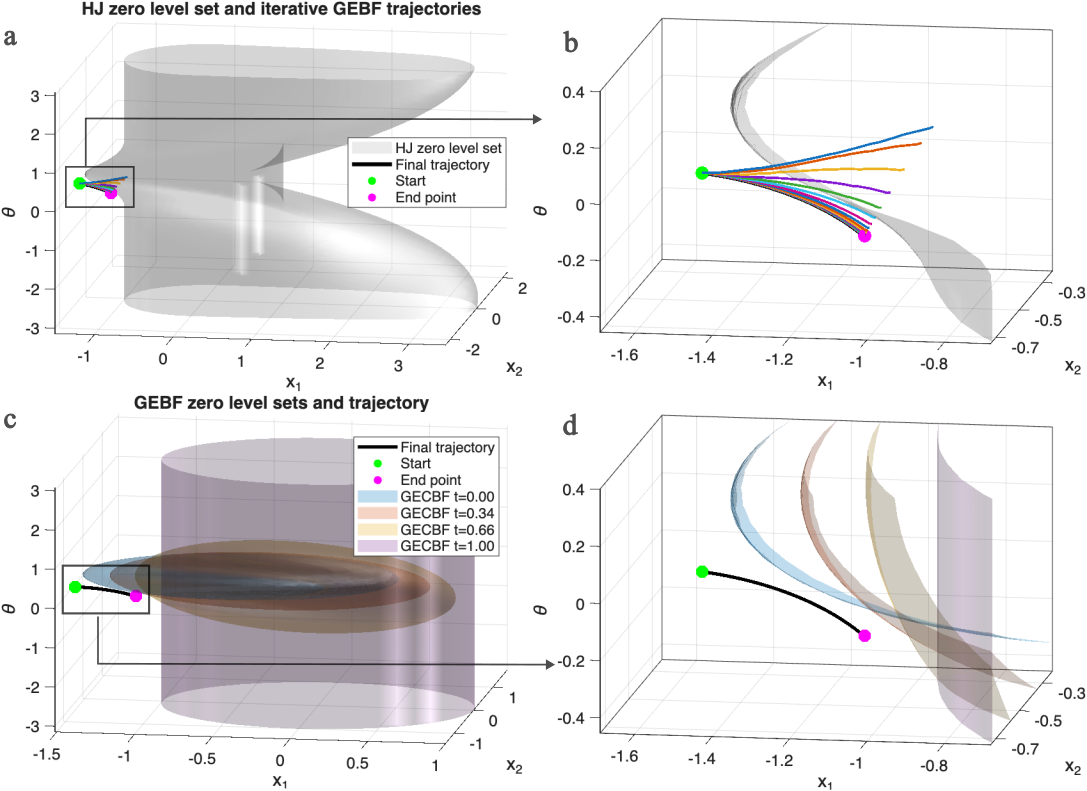}
    \vspace{-0.3cm}
    \caption{Iterative trajectory evolution (a) and its zoomed-in view (b).
Evolution of GEBF zero-level sets (c) and its zoomed-in view (d). At the starting point, $h=0.387$ and $V=0.41$.}
    \vspace{-0.4cm}
    \label{fig:case2_all}
\end{figure}

\vspace{-0.1cm}
\begin{figure}[htb!]
    \centering
    \includegraphics[width=\linewidth]{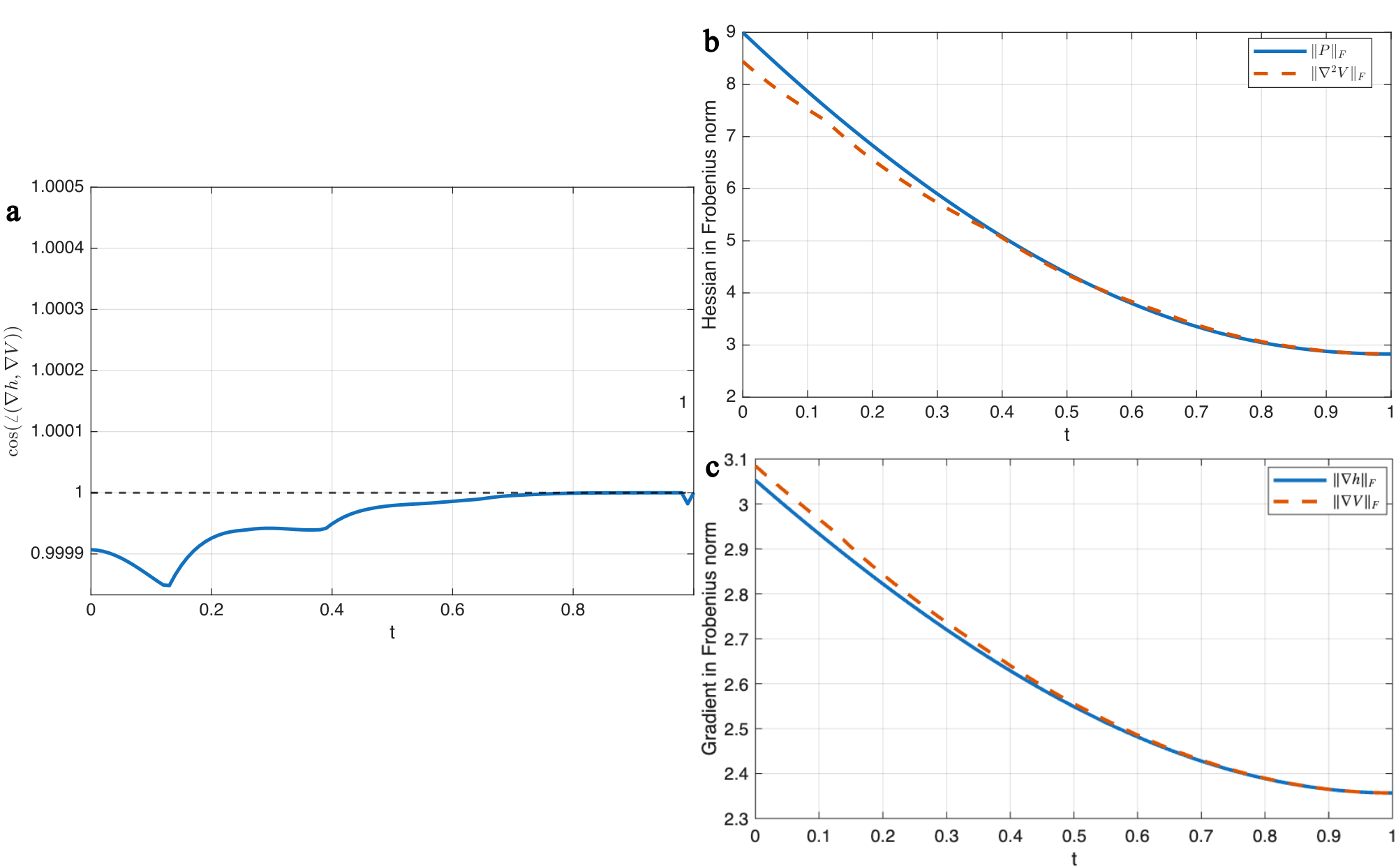}
    \vspace{-0.4cm}
    \caption{
(a) Alignment of the spatial gradients of the GEBF and HJ value function, measured by their cosine similarity
$\cos(\angle(\nabla h,\nabla V))
=
\frac{
\nabla_{\bs z} h(\bs z_{\mathrm{ref}}(t),t)^\top
\nabla_{\bs z} V(\bs z_{\mathrm{ref}}(t),t)
}{
\|\nabla_{\bs z} h(\bs z_{\mathrm{ref}}(t),t)\|_2
\|\nabla_{\bs z} V(\bs z_{\mathrm{ref}}(t),t)\|_2
},$
where $\angle(\nabla h,\nabla V)$ is the angle between the GEBF and HJ gradients.
Comparison of Frobenius norms of the Hessian (b) and Graident (c) of GEBF $h$ and the numerically evaluated HJ value function $V$ along the same trajectory.}
    \vspace{-0.4cm}
    \label{fig:case2_geometry_comparsion}
\end{figure}



\paragraph{Results.}  Figs.~\ref{fig:case2_all}(a)-(c) show the HJ zero-level set together with the trajectories generated during the iterative updates.
The resulting final trajectory remains outside the HJ zero-level set throughout the entire time horizon, indicating that the ego system safely avoids the unsafe region generated by obstacle agent.
Figs.~\ref{fig:case2_all}(c)-(d) show the the evolution of the GEBF zero-level set, which closely matches the HJ zero-level set near the final trajectory at $t=0$.
We further examine an initial state near the zero level set, where the GEBF approximation yields ($h(\boldsymbol z_0,0)\approx-0.0015$), compared with the grid-based HJ reference ($V(\boldsymbol z_0,0)\approx0.035$), as shown in Fig.\ref{fig:case2_value_comparsion_levelset}, Appendix. Despite their proximity to the safety boundary, the results exhibit the one-sided relationship.
Figs.~\ref{fig:case2_geometry_comparsion}(a)-(b) evaluate the local differential geometry captured by the GEBF along the final trajectory. The GEBF gradient remains closely aligned with the HJ gradient, with their cosine similarity remaining near one (i.e., their gradient directions have a strong alignment), while their gradient and Hessian norms exhibit similar trends.
These numerical results provide empirical support for our theoretical findings on first- and second-order HJ geometric consistency.
In addition, the resulting gradient and Hessian components along the trajectory exhibit continuous and smooth variations, providing numerical evidence consistent with Assumption \ref{ass:hj_second_order}, as illustrated in  Fig.~\ref{fig:case2_ac_validation} in Appendix.

Fig.~\ref{fig:case2_rest1}(a) further shows that the approximation error $h-V$ remains small along the trajectory, confirming the local consistency between GEBFs and HJ reachability.
The continuous evolution of the GEBF parameters is in Fig.~\ref{fig:case2_parameters_evolution}, Appendix~\ref{append_experiments_example:three}.
Furthermore, Fig.~\ref{fig:case2_rest1}(b) shows that the GEBF closely approximates the HJ value function for most sampled trajectories, with typical absolute errors on the order of $10^{-2}$.
Larger errors (around $10^{-1}$) occur mainly for larger HJ values. 
Part of this discrepancy is caused by the finite spatial discretization used in helperOC and can be further reduced with a finer computational grid (Fig.~\ref{fig:mean_sample_comparison} in Appendix~\ref{append_experiments_example:three}).


\vspace{-0.2cm}
\begin{figure}[htb!]
    \centering
    \includegraphics[width=\linewidth]{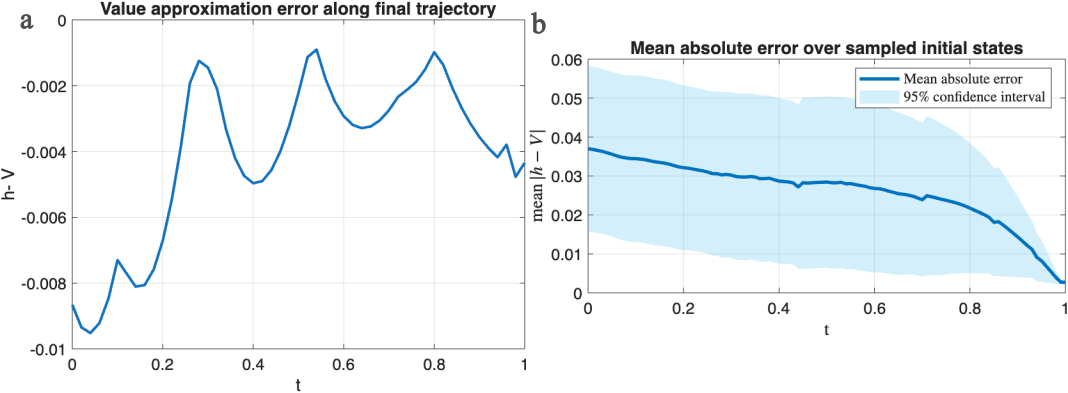}
    \vspace{-0.6cm}
    \caption{(a) Difference between the GEBF and HJ value function at the starting point.
    (b) Mean absolute error $|h-V|$ over $20$ uniformly sampled initial states from $[-1.9,-1.1]\times[-0.9,-0.1]\times[-0.5,0.5]$.}
    \vspace{-0.4cm}
    \label{fig:case2_rest1}
\end{figure}




\subsection{8D Coupled Kinematic Cars: 5D Relative Case}
\label{ex:8d_simple_car}

\paragraph{Environment.} The 8D collision-avoidance problem consists of an ego vehicle and a moving obstacle, both modeled by the 4D kinematic car dynamics (Appendix~\ref{append_ex:8d_simple_car}) 
\cite{rajamani2006vehicle}, a standard benchmark in autonomous driving and motion planning \cite{nachkov2026unlocking,seo2022real}. We adopt the same notation as in the previous example. Unlike the 3D case, the ego and obstacle velocities $v_e$ and $v_o$ are state variables, and the ego control is
$
u=[a_e,\omega_e]^\top\in\mathcal U,
$
where $a_e$ and $\omega_e$ denote acceleration and angular velocity, respectively. The obstacle inputs $a_o$ and $\omega_o$ are assumed to be prescribed.
The complete system has the 8D joint state $(\bs x_e,\bs x_o)$, where $\bs x_e=(p^e,\theta_e,v_e)$ and $\bs x_o=(p^o,\theta_o,v_o)$.
The collision set is $\mathcal L=
\{(\bs x_e,\bs x_o):\|p^o-p^e\|\le R\}.$
Since helperOC is limited to systems of at most five state dimensions, we apply the similar relative-coordinate reduction to obtain the 5D model \eqref{eq:5d_relative_simple_car} (Appendix~\ref{append_ex:8d_simple_car}), where
$
\bs z=(x_1,x_2,\theta,v_e,v_o)^\top
.$
The terminal function is
$
l(\bs z)=x_1^2+x_2^2-R^2.
$
In the simulation, we uniformly sample 20 initial states from $[-2.35,-1.65]\times [-1.15,-0.45]\times [-0.30,0.30]\times [0.85,1.15]\times [0.85,1.15].$

\begin{figure}[htb!]
    \centering
    \includegraphics[width=\linewidth]{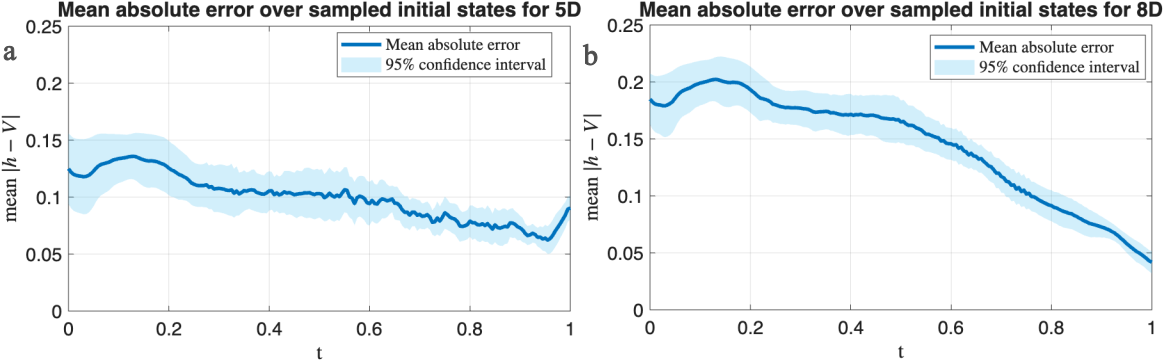}
    \vspace{-0.6cm}
    \caption{Mean absolute approximation error $|h-V|$ for reduced 5D (a) and 8D (b) case over $20$ sampled initial states.}
    \vspace{-0.6cm}
    \label{fig:sample_mean_lowquality_58D}
\end{figure}

\paragraph{Results.}
Fig.~\ref{fig:sample_mean_lowquality_58D} compares the Monte Carlo approximation errors obtained from the $5$D and coupled $8$D formulations.
In Fig.~\ref{fig:sample_mean_lowquality_58D}(a), the mean approximation error remains below $0.15$ throughout the interval $t\in[0,1]$ and exhibits a generally decreasing trend. This corresponds to less than approximately $1.2\%$ of the minimum HJ value ($13.0344$) in the sampled points, demonstrating that the proposed GEBF accurately approximates the local HJ value function.
To further evaluate the scalability, we additionally consider the coupled 8D system described by~\eqref{eq:coupledDynamics}. 
As direct HJ analysis is computationally intractable, the reference value is obtained from the equivalent 5D HJ solution after the relative-coordinate reduction.
Thus, the reported error in Fig.~\ref{fig:sample_mean_lowquality_58D}(b) reflects not only the GEBF approximation but also the discrepancy introduced by the coordinate reduction and nonlinear state transformation.
Even so, the error remains of the same order of magnitude and decreases steadily over time, indicating that GEBFs remain effective for higher-dimensional coupled systems despite the absence of direct HJ solutions.

\vspace{-0.2cm}



\paragraph{Discussion.}
These case studies demonstrate that the proposed framework constructs GEBFs that closely approximate the local HJ value function while generating safe trajectories through iterative geometry evolution. The approximation error remains small across all examples despite increasing system dimension, indicating that the proposed local construction effectively preserves the underlying safety geometry. Table~\ref{tab:comparison} further shows that GEBF evaluation is computationally inexpensive, enabling efficient local HJ safety evaluation in high-dimensional systems.
To further assess scalability, we consider an additional 8D system that does not admit the relative-coordinate reduction used above (Appendix~\ref{append_ex:8d_simple_car_noRela}), for which GEBF construction takes only 1.09 seconds, as reported in Table 1.

\vspace{-0.2cm}
\begin{table}[htb!]
\centering
\scriptsize
\setlength{\tabcolsep}{3pt}
\caption{Computation time comparison (seconds).}
\vspace{-0.4cm}
\label{tab:comparison}
\begin{tabular}{lcccc}
\toprule
Method
& 2D Case
& 3D Relative Case
& 8D Coupled Case
& Irreducible 8D Case \\
&
&
& (5D Relative)
& (in Appendix~\ref{append_ex:8d_simple_car_noRela})  \\
\midrule
helperOC & $3.4$  & $279.12$  & $2879.75$ & N/A \\
Proposed GEBF   & $0.05$  & $0.11$  & $0.38$ & $1.09$  \\
\bottomrule
\end{tabular}
\vspace{-0.4cm}
\end{table}

\begin{remark}
The relative-coordinate reduction is exact only when both the system dynamics and safety specification depend solely on the ego-obstacle relative configuration, so that the relative dynamics remain closed.
If the obstacle dynamics depend on its absolute state (e.g., $\dot v_o=\alpha\tanh(p_x^o)$; see Appendix~\ref{append_ex:8d_simple_car_noRela} for model details and simulation results), or the safety specification depends on absolute positions, the relative dynamics are no longer closed.
Consequently, relative-coordinate formulations are generally inapplicable when the system dynamics depend on absolute states.

\end{remark}

\section{Conclusion}




This paper introduced a GEBF framework for safety evaluation in dynamic environments. 
Rather than computing a global HJ solution, the proposed framework constructs an explicit, time-varying quadratic barrier by propagating local HJ safety geometry along a reference trajectory. Through an iterative coupling of geometric propagation and trajectory refinement, the framework captures the local evolution of safety boundaries and the associated control behavior.

Theoretical analysis established the zeroth-, first-, and second-order geometric consistency of GEBFs with the HJ value function along HJ-optimal characteristics under the stated regularity conditions. 
Furthermore, we established the fixed-point consistency of the trajectory–geometry iteration and a trajectory-wise one-sided relation between the constructed GEBF and the HJ value function, providing a conditional soundness guarantee for safety classification along the reference trajectory. 
Numerical experiments demonstrated accurate approximation of HJ safety geometry, stable iterative refinement across different initializations, and computational efficiency in higher-dimensional dynamical systems.

Future work will investigate the convergence of the proposed trajectory-geometry iteration algorithm, including sufficient conditions for convergence to an HJ-optimal trajectory. 
We will aim to relax the current assumptions by establishing sufficient conditions for second-order regularity of HJ value functions along optimal characteristics, and investigate the propagation of singularities to develop geometric representations capable of capturing nonsmooth behavior. 
These theoretical developments will provide a foundation for extending GEBFs to adversarial multi-agent systems, where competing strategies and switching structures may give rise to more complex HJ–Isaacs value-function geometries.


\section*{Ethics and Privacy Statement}
This work advances the safety evaluation of autonomous systems in dynamic environments by developing computationally efficient, trajectory-based safety representations, with potential benefits for safety-critical applications.
However, the theoretical guarantees rely on specific regularity assumptions, and deploying the proposed method in real-world systems without adequately validating these conditions could lead to inaccurate safety assessments and unintended risks.

\bibliographystyle{ACM-Reference-Format}
\bibliography{AuthorKit27/aaai2027}

@article{crandall1992user,
  title={User’s guide to viscosity solutions of second order partial differential equations},
  author={Crandall, Michael G and Ishii, Hitoshi and Lions, Pierre-Louis},
  journal={Bulletin of the American mathematical society},
  volume={27},
  number={1},
  pages={1--67},
  year={1992}
}

@article{aleksandorov1939almost,
  title={Almost everywhere existence of the second differential of a convex function and some properties of convex functions},
  author={Aleksandorov, AD},
  journal={Leningrad Univ. Ann.},
  volume={37},
  pages={3--35},
  year={1939}
}

@article{ambrosio1993propagation,
  title={On the propagation of singularities of semi-convex functions},
  author={Ambrosio, Luigi and Cannarsa, Piermarco and Soner, H Mete},
  journal={Annali della Scuola Normale Superiore di Pisa-Classe di Scienze},
  volume={20},
  number={4},
  pages={597--616},
  year={1993}
}

@inproceedings{ames2019control,
  title={Control barrier functions: Theory and applications},
  author={Ames, Aaron D and Coogan, Samuel and Egerstedt, Magnus and Notomista, Gennaro and Sreenath, Koushil and Tabuada, Paulo},
  booktitle={2019 18th European control conference (ECC)},
  pages={3420--3431},
  year={2019},
  organization={Ieee}
}

@article{mitchell2005time,
  title={A time-dependent Hamilton-Jacobi formulation of reachable sets for continuous dynamic games},
  author={Mitchell, Ian M and Bayen, Alexandre M and Tomlin, Claire J},
  journal={IEEE Transactions on automatic control},
  volume={50},
  number={7},
  pages={947--957},
  year={2005},
  publisher={IEEE}
}

@book{evans2022partial,
  title={Partial differential equations},
  author={Evans, Lawrence C},
  volume={19},
  year={2022},
  publisher={American mathematical society}
}

@book{federer2014geometric,
  title={Geometric measure theory},
  author={Federer, Herbert},
  year={2014},
  publisher={Springer}
}

@article{wieland2007constructive,
  title={Constructive safety using control barrier functions},
  author={Wieland, Peter and Allg{\"o}wer, Frank},
  journal={IFAC Proceedings Volumes},
  volume={40},
  number={12},
  pages={462--467},
  year={2007},
  publisher={Elsevier}
}

@inproceedings{bansal2017hamilton,
  title={Hamilton-jacobi reachability: A brief overview and recent advances},
  author={Bansal, Somil and Chen, Mo and Herbert, Sylvia and Tomlin, Claire J},
  booktitle={2017 IEEE 56th annual conference on decision and control (CDC)},
  pages={2242--2253},
  year={2017},
  organization={IEEE}
}

@article{chen2018hamilton,
  title={Hamilton--Jacobi reachability: Some recent theoretical advances and applications in unmanned airspace management},
  author={Chen, Mo and Tomlin, Claire J},
  journal={Annual Review of Control, Robotics, and Autonomous Systems},
  volume={1},
  number={1},
  pages={333--358},
  year={2018},
  publisher={Annual Reviews}
}

@article{ganai2024hamilton,
  title={Hamilton-jacobi reachability in reinforcement learning: A survey},
  author={Ganai, Milan and Gao, Sicun and Herbert, Sylvia L},
  journal={IEEE Open Journal of Control Systems},
  volume={3},
  pages={310--324},
  year={2024},
  publisher={IEEE}
}

@book{bardi1997optimal,
  title={Optimal control and viscosity solutions of Hamilton-Jacobi-Bellman equations},
  author={Bardi, Martino and Dolcetta, Italo Capuzzo and others},
  volume={12},
  year={1997},
  publisher={Springer}
}

@inproceedings{mitchell2005toolbox,
  title={A toolbox of Hamilton-Jacobi solvers for analysis of nondeterministic continuous and hybrid systems},
  author={Mitchell, Ian M and Templeton, Jeremy A},
  booktitle={International workshop on hybrid systems: computation and control},
  pages={480--494},
  year={2005},
  organization={Springer}
}

@inproceedings{fisac2015reach,
  title={Reach-avoid problems with time-varying dynamics, targets and constraints},
  author={Fisac, Jaime F and Chen, Mo and Tomlin, Claire J and Sastry, S Shankar},
  booktitle={Proceedings of the 18th international conference on hybrid systems: computation and control},
  pages={11--20},
  year={2015}
}

@article{ames2016control,
  title={Control barrier function based quadratic programs for safety critical systems},
  author={Ames, Aaron D and Xu, Xiangru and Grizzle, Jessy W and Tabuada, Paulo},
  journal={IEEE Transactions on Automatic Control},
  volume={62},
  number={8},
  pages={3861--3876},
  year={2016},
  publisher={IEEE}
}

@inproceedings{yu2023sequential,
  title={Sequential neural barriers for scalable dynamic obstacle avoidance},
  author={Yu, Hongzhan and Hirayama, Chiaki and Yu, Chenning and Herbert, Sylvia and Gao, Sicun},
  booktitle={2023 IEEE/RSJ International Conference on Intelligent Robots and Systems (IROS)},
  pages={11241--11248},
  year={2023},
  organization={IEEE}
}

@inproceedings{chen2025int2planner,
  title={Int2Planner: An intention-based multi-modal motion planner for integrated prediction and planning},
  author={Chen, Xiaolei and Yan, Junchi and Liao, Wenlong and He, Tao and Peng, Pai},
  booktitle={Proceedings of the AAAI Conference on Artificial Intelligence},
  volume={39},
  number={14},
  pages={14558--14566},
  year={2025}
  }

@book{jurdjevic1997geometric,
  title={Geometric control theory},
  author={Jurdjevic, Velimir},
  year={1997},
  publisher={Cambridge university press}
}

@inproceedings{Tahmid2025ValueIteration,
  author    = {Sheikh A. Tahmid and Gennaro Notomista},
  title     = {Value Iteration for Learning Concurrently Executable Robotic Control Tasks},
  booktitle = {Proceedings of the 24th International Conference on Autonomous Agents and Multiagent Systems (AAMAS)},
  year      = {2025},
  pages     = {2006--2014},
  publisher = {International Foundation for Autonomous Agents and Multiagent Systems (IFAAMAS)},
  address   = {Detroit, MI, USA},
  isbn      = {979-8-4007-1426-9}
}

@article{evans1989hamiltonian,
  title={The Hamiltonian--Jacobi--Bellman equation for time-optimal control},
  author={Evans, LC and James, MR},
  journal={SIAM journal on control and optimization},
  volume={27},
  number={6},
  pages={1477--1489},
  year={1989},
  publisher={SIAM}
}

@book{rajamani2006vehicle,
  title={Vehicle dynamics and control},
  author={Rajamani, Rajesh},
  year={2006},
  publisher={Springer}
}

@inproceedings{nachkov2026unlocking,
  title={Unlocking efficient vehicle dynamics modeling via analytic world models},
  author={Nachkov, Asen and Paudel, Danda Pani and Zaech, Jan-Nico and Scaramuzza, Davide and Van Gool, Luc},
  booktitle={Proceedings of the AAAI Conference on Artificial Intelligence},
  volume={40},
  number={29},
  pages={24467--24475},
  year={2026}
}

@article{seo2022real,
  title={Real-time robust receding horizon planning using Hamilton--Jacobi reachability analysis},
  author={Seo, Hoseong and Lee, Donggun and Son, Clark Youngdong and Jang, Inkyu and Tomlin, Claire J and Kim, H Jin},
  journal={IEEE Transactions on Robotics},
  volume={39},
  number={1},
  pages={90--109},
  year={2022},
  publisher={IEEE}
}

@book{strogatz2024nonlinear,
  title={Nonlinear dynamics and chaos: with applications to physics, biology, chemistry, and engineering},
  author={Strogatz, Steven H},
  year={2024},
  publisher={Chapman and Hall/CRC}
}

@article{amodei2016concrete,
  title={Concrete problems in AI safety},
  author={Amodei, Dario and Olah, Chris and Steinhardt, Jacob and Christiano, Paul and Schulman, John and Man{\'e}, Dan},
  journal={arXiv preprint arXiv:1606.06565},
  year={2016}
}

@article{paden2016survey,
  title={A survey of motion planning and control techniques for self-driving urban vehicles},
  author={Paden, Brian and {\v{C}}{\'a}p, Michal and Yong, Sze Zheng and Yershov, Dmitry and Frazzoli, Emilio},
  journal={IEEE Transactions on intelligent vehicles},
  volume={1},
  number={1},
  pages={33--55},
  year={2016},
  publisher={IEEE}
}

@article{hsu2023safety,
  title={The safety filter: A unified view of safety-critical control in autonomous systems},
  author={Hsu, Kai-Chieh and Hu, Haimin and Fisac, Jaime F},
  journal={Annual Review of Control, Robotics, and Autonomous Systems},
  volume={7},
  year={2023},
  publisher={Annual Reviews}
}

@inproceedings{ma2024conformal,
  title     = {Conformal Reachability for Safe Control in Unknown Environments},
  author    = {Ma, Xinhang and Wu, Junlin and Kantaros, Yiannis and Vorobeychik, Yevgeniy},
  booktitle = {Proceedings of the 25th International Conference on Autonomous Agents and Multiagent Systems },
  pages     = {192--201},
  year      = {2026},
  publisher = {AAMAS}
}

@inproceedings{luo2023towards,
  title={Towards safe and resilient autonomy in multi-robot systems},
  author={Luo, Wenhao},
  booktitle={Proceedings of the AAAI Conference on Artificial Intelligence},
  volume={37},
  number={13},
  pages={15449--15449},
  year={2023}
}

@article{garg2024learning,
  title={Learning safe control for multi-robot systems: Methods, verification, and open challenges},
  author={Garg, Kunal and Zhang, Songyuan and So, Oswin and Dawson, Charles and Fan, Chuchu},
  journal={Annual Reviews in Control},
  volume={57},
  pages={100948},
  year={2024},
  publisher={Elsevier}
}

@article{darbon2016algorithms,
  title={Algorithms for overcoming the curse of dimensionality for certain Hamilton--Jacobi equations arising in control theory and elsewhere},
  author={Darbon, J{\'e}r{\^o}me and Osher, Stanley},
  journal={Research in the Mathematical Sciences},
  volume={3},
  number={1},
  pages={19},
  year={2016},
  publisher={Springer}
}

@article{osher2004level,
  title={Level set methods and dynamic implicit surfaces},
  author={Osher, Stanley and Fedkiw, Ronald and Piechor, Krzysztof},
  journal={Appl. Mech. Rev.},
  volume={57},
  number={3},
  pages={B15--B15},
  year={2004}
}

@inproceedings{chen2017exact,
  title     = {Exact and efficient Hamilton-Jacobi guaranteed safety analysis via system decomposition},
  author    = {Chen, Mo and Herbert, Sylvia and Tomlin, Claire J.},
  booktitle = {2017 IEEE International Conference on Robotics and Automation (ICRA)},
  pages     = {87--92},
  year      = {2017},
  organization = {IEEE}
}

@inproceedings{bansal2021deepreach,
  title={Deepreach: A deep learning approach to high-dimensional reachability},
  author={Bansal, Somil and Tomlin, Claire J},
  booktitle={2021 IEEE International Conference on Robotics and Automation (ICRA)},
  pages={1817--1824},
  year={2021},
  organization={IEEE}
}

@article{yang2025scalable,
  title={Scalable synthesis of formally verified neural value function for hamilton-jacobi reachability analysis},
  author={Yang, Yujie and Hu, Hanjiang and Wei, Tianhao and Li, Shengbo Eben and Liu, Changliu},
  journal={Journal of Artificial Intelligence Research},
  volume={83},
  year={2025}
}

@inproceedings{herbert2021scalable,
  title={Scalable learning of safety guarantees for autonomous systems using hamilton-jacobi reachability},
  author={Herbert, Sylvia and Choi, Jason J and Sanjeev, Suvansh and Gibson, Marsalis and Sreenath, Koushil and Tomlin, Claire J},
  booktitle={2021 IEEE International Conference on Robotics and Automation (ICRA)},
  pages={5914--5920},
  year={2021},
  organization={IEEE}
}

@article{zhang2023exact,
  title={Exact verification of relu neural control barrier functions},
  author={Zhang, Hongchao and Wu, Junlin and Vorobeychik, Yevgeniy and Clark, Andrew},
  journal={Advances in neural information processing systems},
  volume={36},
  pages={5685--5705},
  year={2023}
}

@article{chen2018decomposition,
  title={Decomposition of reachable sets and tubes for a class of nonlinear systems},
  author={Chen, Mo and Herbert, Sylvia L and Vashishtha, Mahesh S and Bansal, Somil and Tomlin, Claire J},
  journal={IEEE Transactions on Automatic Control},
  volume={63},
  number={11},
  pages={3675--3688},
  year={2018},
  publisher={IEEE}
}

@inproceedings{chen2021backup,
  title={Backup control barrier functions: Formulation and comparative study},
  author={Chen, Yuxiao and Jankovic, Mrdjan and Santillo, Mario and Ames, Aaron D},
  booktitle={2021 60th IEEE Conference on Decision and Control (CDC)},
  pages={6835--6841},
  year={2021},
  organization={IEEE}
}

@inproceedings{cheng2019end,
  title={End-to-end safe reinforcement learning through barrier functions for safety-critical continuous control tasks},
  author={Cheng, Richard and Orosz, G{\'a}bor and Murray, Richard M and Burdick, Joel W},
  booktitle={Proceedings of the AAAI conference on artificial intelligence},
  volume={33},
  number={01},
  pages={3387--3395},
  year={2019}
}

@inproceedings{taylor2020learning,
  title={Learning for safety-critical control with control barrier functions},
  author={Taylor, Andrew and Singletary, Andrew and Yue, Yisong and Ames, Aaron},
  booktitle={Learning for dynamics and control},
  pages={708--717},
  year={2020},
  organization={PMLR}
}

@article{igarashi2019time,
  title={Time-varying control barrier function and its application to environment-adaptive human assist control},
  author={Igarashi, Motoi and Tezuka, Issei and Nakamura, Hisakazu},
  journal={IFAC-PapersOnLine},
  volume={52},
  number={16},
  pages={735--740},
  year={2019},
  publisher={Elsevier}
}

@article{dai2025differentiable,
  title={Differentiable optimization based time-varying control barrier functions for dynamic obstacle avoidance},
  author={Dai, Bolun and Khorrambakht, Rooholla and Krishnamurthy, Prashanth and Khorrami, Farshad},
  journal={Robotics and Autonomous Systems},
  pages={105182},
  year={2025},
  publisher={Elsevier}
}

@article{luo2021learning,
  title={Learning barrier certificates: Towards safe reinforcement learning with zero training-time violations},
  author={Luo, Yuping and Ma, Tengyu},
  journal={Advances in Neural Information Processing Systems},
  volume={34},
  pages={25621--25632},
  year={2021}
}

@inproceedings{belardinelli2025probabilistic,
  title={Probabilistic shielding for safe reinforcement learning},
  author={Belardinelli, Francesco and Goodall, Alexander W and others},
  booktitle={Proceedings of the AAAI Conference on Artificial Intelligence},
  volume={39},
  number={15},
  pages={16091--16099},
  year={2025}
}

@article{jin2020neural,
  title={Neural certificates for safe control policies},
  author={Jin, Wanxin and Wang, Zhaoran and Yang, Zhuoran and Mou, Shaoshuai},
  journal={arXiv preprint arXiv:2006.08465},
  year={2020}
}

@article{dawson2023safe,
  title={Safe control with learned certificates: A survey of neural lyapunov, barrier, and contraction methods for robotics and control},
  author={Dawson, Charles and Gao, Sicun and Fan, Chuchu},
  journal={IEEE Transactions on Robotics},
  volume={39},
  number={3},
  pages={1749--1767},
  year={2023},
  publisher={IEEE}
}

@article{sun2023conformal,
  title={Conformal prediction for uncertainty-aware planning with diffusion dynamics model},
  author={Sun, Jiankai and Jiang, Yiqi and Qiu, Jianing and Nobel, Parth and Kochenderfer, Mykel J and Schwager, Mac},
  journal={Advances in Neural Information Processing Systems},
  volume={36},
  pages={80324--80337},
  year={2023}
}

@inproceedings{choi2021robust,
  title={Robust control barrier--value functions for safety-critical control},
  author={Choi, Jason J and Lee, Donggun and Sreenath, Koushil and Tomlin, Claire J and Herbert, Sylvia L},
  booktitle={2021 60th IEEE Conference on Decision and Control (CDC)},
  pages={6814--6821},
  year={2021},
  organization={IEEE}
}

@article{hirsch2025viscosity,
  title={Viscosity CBFs: Bridging the Control Barrier Function and Hamilton-Jacobi Reachability Frameworks in Safe Control Theory},
  author={Hirsch, Dylan and Fisac, Jaime Fern{\'a}ndez and Herbert, Sylvia},
  journal={arXiv preprint arXiv:2510.09929},
  year={2025}
}

@article{yegorov2021perspectives,
  title={Perspectives on characteristics based curse-of-dimensionality-free numerical approaches for solving Hamilton--Jacobi equations},
  author={Yegorov, Ivan and Dower, Peter M},
  journal={Applied Mathematics \& Optimization},
  volume={83},
  number={1},
  pages={1--49},
  year={2021},
  publisher={Springer}
}

@article{chow2019algorithm,
  title={Algorithm for overcoming the curse of dimensionality for state-dependent Hamilton-Jacobi equations},
  author={Chow, Yat Tin and Darbon, J{\'e}r{\^o}me and Osher, Stanley and Yin, Wotao},
  journal={Journal of Computational Physics},
  volume={387},
  pages={376--409},
  year={2019},
  publisher={Elsevier}
}

@article{subbotina2006method,
  title={The method of characteristics for Hamilton—Jacobi equations and applications to dynamical optimization},
  author={Subbotina, NN},
  journal={Journal of mathematical sciences},
  volume={135},
  number={3},
  pages={2955--3091},
  year={2006},
  publisher={Springer}
}

@inproceedings{tonkens2024patching,
  title={Patching approximately safe value functions leveraging local hamilton-jacobi reachability analysis},
  author={Tonkens, Sander and Toofanian, Alex and Qin, Zhizhen and Gao, Sicun and Herbert, Sylvia},
  booktitle={2024 IEEE 63rd Conference on Decision and Control (CDC)},
  pages={3577--3584},
  year={2024},
  organization={IEEE}
}

@article{sharpless2026conservative,
  title={Conservative linear envelopes for nonlinear, high-dimensional, hamilton-jacobi reachability},
  author={Sharpless, William and Chow, Yat Tin and Herbert, Sylvia},
  journal={IEEE Transactions on Automatic Control},
  year={2026},
  publisher={IEEE}
}

@article{kumar2023fast,
  title={Fast, smooth, and safe: implicit control barrier functions through reach-avoid differential dynamic programming},
  author={Kumar, Athindran Ramesh and Hsu, Kai-Chieh and Ramadge, Peter J and Fisac, Jaime F},
  journal={IEEE Control Systems Letters},
  volume={7},
  pages={2994--2999},
  year={2023},
  publisher={IEEE}
}

@article{tedrake2010lqr,
  title={LQR-trees: Feedback motion planning via sums-of-squares verification},
  author={Tedrake, Russ and Manchester, Ian R and Tobenkin, Mark and Roberts, John W},
  journal={The International Journal of Robotics Research},
  volume={29},
  number={8},
  pages={1038--1052},
  year={2010},
  publisher={SAGE Publications Sage UK: London, England}
}

@article{majumdar2017funnel,
  title={Funnel libraries for real-time robust feedback motion planning},
  author={Majumdar, Anirudha and Tedrake, Russ},
  journal={The International Journal of Robotics Research},
  volume={36},
  number={8},
  pages={947--982},
  year={2017},
  publisher={SAGE Publications Sage UK: London, England}
}

@article{cannarsa2015second,
  title={Second-order sensitivity relations and regularity of the value function for Mayer's problem in optimal control},
  author={Cannarsa, Piermarco and Frankowska, H{\'e}l{\`e}ne and Scarinci, Teresa},
  journal={SIAM Journal on Control and Optimization},
  volume={53},
  number={6},
  pages={3642--3672},
  year={2015},
  publisher={SIAM}
}

@article{feng2025bridging,
  title={Bridging model predictive control and deep learning for scalable reachability analysis},
  author={Feng, Zeyuan and Qiu, Le and Bansal, Somil},
  journal={arXiv preprint arXiv:2505.03830},
  year={2025}
}

@article{borquez2024safety,
  title={On safety and liveness filtering using hamilton--jacobi reachability analysis},
  author={Borquez, Javier and Chakraborty, Kaustav and Wang, Hao and Bansal, Somil},
  journal={IEEE Transactions on Robotics},
  volume={40},
  pages={4235--4251},
  year={2024},
  publisher={IEEE}
}

@book{royden1988real,
  title={Real analysis},
  author={Royden, Halsey Lawrence and Fitzpatrick, Patrick},
  volume={32},
  year={1988},
  publisher={Macmillan, New York}
}



\appendix


\newpage

~\\

\newpage

\noindent\textbf{\Large{APPENDIX}}
\par\smallskip

\section{Supplementary of Section~\ref{sec_viscosity_sup}}\label{append_sec_viscosity_sup}


\paragraph{Proof of Lemma~\ref{iff_safety}:}

\begin{proof}
The HJ variational inequality characterization follows from the standard dynamic programming formulation of
finite-horizon reachability; see, e.g.,
\cite{choi2021robust,fisac2015reach}.
In particular, the dynamic programming principle
accounts for the minimum safety level attained
over the time horizon, leading to the obstacle
term $l(\bs z)-V$ in
Eq.~\eqref{eq:HJVI}.

To establish the safe-set characterization,
observe that, under attainment of the optimal
value,
\begin{align*}
&V(\bs z_0,0)\geq 0
\nonumber\\
&\quad\Longleftrightarrow
\exists u(\cdot)\in\mathcal U_{[0,T]}
\text{ such that }
\inf_{s\in[0,T]}
l(\bs z_u(s))\geq 0
\nonumber\\
&\quad\Longleftrightarrow
\exists u(\cdot)\in\mathcal U_{[0,T]}
\text{ such that }
l(\bs z_u(s))\geq 0,
\quad \forall s\in[0,T].
\end{align*}
\end{proof}

\begin{lemma}[Alexandrov Theorem~{\cite{aleksandorov1939almost}}]
\label{prop:alexandrov}
If $\kappa:\Omega\to\mathbb R$ is semiconvex, then $\kappa$ is twice differentiable
almost everywhere in $\Omega$.
\end{lemma}

\begin{definition}[Second-order Superjet~{\cite{crandall1992user}}]
Let $\kappa:\Omega\to\mathbb R$ be upper semicontinuous. The second-order superjet
of $\kappa$ at $\bs z\in\Omega$ is defined as
\begin{align*}
    &J^{2,+}\kappa(\bs z)
=
\bigg\{
(\bs p,\bs X)\in\mathbb R^{2n}\times\mathbb S^{2n}:\kappa(\bs y)
\le
\kappa(\bs z)
\\
&+
\bs p^\top(\bs y-\bs z)
+
\frac12(\bs y-\bs z)^\top\bs X(\bs y-\bs z)
+
o(\|\bs y-\bs z\|^2)
\bigg\},
\end{align*}
where $\mathbb S^{2n}$ is the set of symmetric $2n\times 2n$ matrices.
The second-order subjet $J^{2,-}\kappa(\bs z)$ is defined analogously by reversing
the inequality.
\end{definition}



\paragraph{Proof of Proposition~\ref{propo_cbf_supersolution}:}
\begin{proof}
Since \(h\in C^1(\Omega\times(0,T))\), it is enough to verify the
supersolution inequality in the classical sense. For any
\((\bs z,t)\in\Omega\times(0,T)\), Eq.~\eqref{eq:hj_type_barrier_ineq}
implies
\begin{equation}
    \frac{\partial h}{\partial t}(\bs z,t)
    +
    H(\bs z,\nabla_{\bs z}h(\bs z,t))
    \ge 0 .
\end{equation}
Therefore, if \(\varphi\in C^\infty(\Omega\times(0,T))\) touches \(h\) from
below at \((\bs z^*,t^*)\), then
\[
    \nabla_{\bs z}\varphi(\bs z^*,t^*)
    =
    \nabla_{\bs z}h(\bs z^*,t^*),
    \qquad
    \frac{\partial \varphi}{\partial t}(\bs z^*,t^*)
    =
    \frac{\partial h}{\partial t}(\bs z^*,t^*).
\]
Substituting these identities into the above inequality yields
\[
    \frac{\partial \varphi}{\partial t}(\bs z^*,t^*)
    +
    H\!\left(
        \bs z^*,
        \nabla_{\bs z}\varphi(\bs z^*,t^*)
    \right)
    \ge 0 .
\]
Hence \(h\) is a viscosity supersolution of
Eq.~\eqref{eq:hj_continuation}.
\end{proof}

\paragraph{Proof of Proposition~\ref{pro:local_second_order_relation}:}
\begin{proof}
Since $\Psi_\theta$ attains a local maximum at
$(\bs z_1^\theta,\bs z_2^\theta)$, the doubling-of-variables argument and the
Crandall--Ishii Lemma \cite{crandall1992user} imply that there exist matrices
$\bs X^\theta,\bs Y^\theta\in\mathbb S^{2n}$ such that
\[
\left(
\theta(\bs z_1^\theta-\bs z_2^\theta),
\bs X^\theta
\right)
\in
J^{2,+}V(\bs z_1^\theta,t),
\]
and
\[
\left(
\theta(\bs z_1^\theta-\bs z_2^\theta),
\bs Y^\theta
\right)
\in
J^{2,-}h(\bs z_2^\theta,t).
\]
The matrix inequality in the Crandall--Ishii lemma gives, in particular along
identical perturbation directions, that
\[
\bs X^\theta
\preceq
\bs Y^\theta .
\]
Since $h(\cdot,t)$ is $C^2$, the structure of the second-order subjet gives
\[
\bs Y^\theta
\preceq
\nabla_{\bs z}^{2}h(\bs z_2^\theta,t).
\]
Therefore,
\[
\bs X^\theta
\preceq
\nabla_{\bs z}^{2}h(\bs z_2^\theta,t).
\]

If $V(\cdot,t)$ is twice differentiable at $\bs z_1^\theta$, then its
second-order superjet is compatible with the classical Hessian. In particular,
\[
\nabla_{\bs z}^{2}V(\bs z_1^\theta,t)
\preceq
\bs X^\theta .
\]
Combining this with the previous inequality yields
\[
\nabla_{\bs z}^{2}V(\bs z_1^\theta,t)
\preceq
\nabla_{\bs z}^{2}h(\bs z_2^\theta,t).
\]
This proves the desired local second-order geometric relation.
\end{proof}

\begin{proposition}[Distance Estimate for Contact Points]\label{rela_max_points}
    Suppose value function $V\in \mathcal{C}(\Omega\times (0,\infty))$ is  Lipschitz continuous and the CBF $h\in \mathcal{C}^{2}(\Omega\times (0,\infty))$ satisfies that $\exists \epsilon>0$
\[
0 \le h(\bs z,t)-V(\bs z,t) \le \epsilon.
\]
Fix $t\in(0,T)$ and assume that
$V(\cdot,t)$ is semiconvex on $\Omega$.
For $\alpha>0$, define
\[
\Psi(\bs z_1,\bs z_2)
=
V(\bs z_1,t)-h(\bs z_2,t)-\frac{\alpha}{2}\|\bs z_1-\bs z_2\|^2.
\]
Let $(\bs z^{\theta}_{1},\bs z^{\theta}_{2})$ be a maximizer of $\Psi$. Then
\[
\|\bs z^{\theta}_{1}-\bs z^{\theta}_{2}\|
\le
\min \left\{ \frac{2L_h}{\alpha},\frac{L+\sqrt{L^2+2\alpha\varepsilon}}{\alpha} \right\},
\]
where $L_h$ and $L$ is the Lipschitz constant of the CBF $h$ and value function $V.$ 
\end{proposition}
\begin{proof}
    From the maximality of $(\bs z^{\theta}_{1},\bs z^{\theta}_{2})$, we have
\[
\Psi(\bs z^{\theta}_{1},\bs z^{\theta}_{2})
\ge
\Psi(\bs z^{\theta}_{1},\bs z^{\theta}_{1}) \text{ and }
\Psi(\bs z^{\theta}_{1},\bs z^{\theta}_{2})
\ge
\Psi(\bs z^{\theta}_{2},\bs z^{\theta}_{2}).
\]
Thus,
\[
V(\bs z^{\theta}_{1},t)-h(\bs z^{\theta}_{2},t)-\frac{\alpha}{2}\|\bs z^{\theta}_{1}-\bs z^{\theta}_{2}\|^2
\ge
V(\bs z^{\theta}_{1},t)-h(\bs z^{\theta}_{1},t),
\]
\[
V(\bs z^{\theta}_{1},t)-h(\bs z^{\theta}_{2},t)-\frac{\alpha}{2}\|\bs z^{\theta}_{1}-\bs z^{\theta}_{2}\|^2
\ge
V(\bs z^{\theta}_{2},t)-h(\bs z^{\theta}_{2},t).
\]
Rearranging gives
\[
h(\bs z^{\theta}_{1},t)-h(\bs z^{\theta}_{2},t)
\ge
\frac{\alpha}{2}\|\bs z^{\theta}_{1}-\bs z^{\theta}_{2}\|^2,
\]
\[
V(\bs z^{\theta}_{1},t)-V(\bs z^{\theta}_{2},t)
\ge
\frac{\alpha}{2}\|\bs z^{\theta}_{1}-\bs z^{\theta}_{2}\|^2.
\]
Using the bound $0\le h(\bs z,t)-V(\bs z,t)\le\varepsilon$ and the Lipschitz continuity of $V$, we obtain
\[
h(\bs z^{\theta}_{1},t)-h(\bs z^{\theta}_{2},t)
\le
\varepsilon
+
L\|\bs z^{\theta}_{1}-\bs z^{\theta}_{2}\|
\]
and 
\[
V(\bs z^{\theta}_{1},t)-V(\bs z^{\theta}_{2})
\le
h(\bs z^{\theta}_{1})-h(\bs z^{\theta}_{2}).
\]
Combining the inequalities yields
\[
\frac{\alpha}{2}|\bs z^{\theta}_{1}-\bs z^{\theta}_{2}|^2
\le
\varepsilon
+
L\|\bs z^{\theta}_{1}-\bs z^{\theta}_{2}\|.
\]
and 
\[
V(\bs z^{\theta}_{1})-V(\bs z^{\theta}_{2})
\le
L_h\|\bs z^{\theta}_{1}-\bs z^{\theta}_{2}\|.
\]
Then, we can have 
\[
\|\bs z^{\theta}_{1}-\bs z^{\theta}_{2}\|
\le
\frac{L+\sqrt{L^2+2\alpha\varepsilon}}{\alpha}.
\]
and 
\[
\|\bs z^{\theta}_{1}-\bs z^{\theta}_{2}\|
\le
\frac{2L_h}{\alpha}.
\]

Overall, 
\[
\|\bs z^{\theta}_{1}-\bs z^{\theta}_{2}\|
\le
\min \left\{ \frac{2L_h}{\alpha},\frac{L+\sqrt{L^2+2\alpha\varepsilon}}{\alpha} \right\}.
\]
\end{proof}  
\begin{remark}
    This proposition implies that the distance between the local maximizers $\bs z^{\theta}_{1}$ and $\bs z^{\theta}_{2}$ becomes arbitrarily small as $\alpha \to \infty$
\end{remark}

\section{Supplementary of Section~\ref{sec_gebf_design}}\label{append_sec_main}

\begin{lemma}[Zero-measure Switching Set]
\label{lem:zero_switch_set}
Let
\[
\mathcal S_{\rm sw}
=
\{\bs z\in\Omega:\boldsymbol\phi(\bs z,t)=0\},
\]
where $\boldsymbol\phi$ is a smooth switching function.
If $\boldsymbol\phi$ satisfies Assumption \ref{ass:hamilton_maximizer}, i.e.,
\[
\nabla_{\bs z}\boldsymbol\phi(\bs z,t)\neq 0
\]
on $\mathcal S_{\rm sw}$, then
\[
\mathcal L^{2n}(\mathcal S_{\rm sw})=0.
\]
\end{lemma}
\begin{proof}
    By the implicit function theorem \cite{evans2022partial}, $\mathcal S_{\rm sw}$ is locally the graph of a $C^1$ function from $\mathbb{R}^{2n-1}$ to $\mathbb{R}$ \cite{federer2014geometric}. Hence, it is an $(2n-1)$-dimensional embedded submanifold and has zero Lebesgue measure in $\mathbb{R}^{2n}$.
\end{proof}

\paragraph{Proof of Theorem~\ref{thm:hessian_dynamics_value}:}


\begin{proof}
Let
\[
\bs g(\bs z,t)
=
\nabla_{\bs z}V(\bs z,t),
\qquad
\bs H(\bs z,t)
=
\nabla_{\bs z}^{2}V(\bs z,t).
\]

We first establish the gradient and Hessian
evolution equations locally away from the
switching times, and then extend the resulting
identities to almost every $t\in[0,T]$.

\medskip
\noindent
\textbf{Step 1: Local characteristic flow.}

Let $\Sigma\subset[0,T]$ denote the
switching-time set, and fix an arbitrary
$t_0\in(0,T)\setminus\Sigma$.
By Assumption~\ref{ass:hj_second_order},
there exists a space-time neighborhood of
$(\bs z_{u}^*(t_0),t_0)$ in which
$V\in C^{1,2}$ and the Hamiltonian maximizer
is represented by a fixed bang-control
branch $\bar u$ such that $u^*(t)\equiv\bar u,$ for $(\bs z,t)\in \mathcal{N}$, where $\mathcal{N}$ be an open neighborhood in state–time space of a smooth segment of the optimal characteristic, chosen to avoid the switching set.

Define $\bs b(\bs z)
=
F(\bs z,\bar u).$
Since $F(\cdot,\bar u)\in C^2$, the vector
field $\bs b$ is twice continuously
differentiable.
In this neighborhood, the HJ equation
reduces to the classical transport equation
\begin{equation}
\label{eq:local_hj_transport}
V_t(\bs z,t)
+
\nabla_{\bs z}V(\bs z,t)^\top
\bs b(\bs z)
=
0.
\end{equation}

Let $\Phi_{t,s}(\bs z)$ denote the local
flow generated by $\bs b$, namely
\[
\frac{\partial}{\partial s}
\Phi_{t,s}(\bs z)
=
\bs b(\Phi_{t,s}(\bs z)),
\qquad
\Phi_{t,t}(\bs z)=\bs z, \qquad  t\leq s.
\]

For sufficiently close $t$ and $s$ in
the chosen neighborhood, the chain rule
and Eq.~\eqref{eq:local_hj_transport} give
\[
\begin{aligned}
\frac{d}{ds}
V(\Phi_{t,s}(\bs z),s)
&=
V_t(\Phi_{t,s}(\bs z),s)
+
\nabla_{\bs z}V(\Phi_{t,s}(\bs z),s)^\top
\bs b(\Phi_{t,s}(\bs z))
=0.
\end{aligned}
\]
Thus,
\begin{equation}
\label{eq:value_flow_identity}
V(\bs z,t)
=
V(\Phi_{t,s}(\bs z),s).
\end{equation}

\medskip
\noindent
\textbf{Step 2: First- and second-order
state differentiation.}

Define the first- and second-order
derivatives of the local flow by
\[
\bs A(t,s,\bs z)
=
D_{\bs z}\Phi_{t,s}(\bs z)
\]
and
\[
\bs B_j(t,s,\bs z)
=
D_{\bs z}^{2}\Phi_{t,s,j}(\bs z),
\qquad j=1,\ldots,2n,
\]
where $2n$ denotes the dimension of the
augmented state.

Differentiating
Eq.~\eqref{eq:value_flow_identity}
once with respect to $\bs z$ yields
\begin{equation}
\label{eq:gradient_flow_identity}
\nabla_{\bs z}V(\bs z,t)
=
\bs A(t,s,\bs z)^\top
\nabla_{\bs z}V(\Phi_{t,s}(\bs z),s).
\end{equation}

Differentiating the same identity twice
and applying the second-order chain rule
gives
\begin{equation}
\label{eq:hessian_flow_identity}
\begin{aligned}
\nabla_{\bs z}^{2}V(\bs z,t)
={}&
\bs A^\top
\nabla_{\bs z}^{2}
V(\Phi_{t,s}(\bs z),s)
\bs A
+
\sum_{j=1}^{2n}
\partial_{z_j}V(\Phi_{t,s}(\bs z),s)
\bs B_j.
\end{aligned}
\end{equation}


Along the selected optimal characteristic,
\[
\Phi_{t,s}(\bs z_{u}^*(t))
=
\bs z_{u}^*(s).
\]

For simplicity, write
\[
\bs g(t)
=
\nabla_{\bs z}V(\bs z_{u}^*(t),t),
\qquad
\bs H(t)
=
\nabla_{\bs z}^{2}V(\bs z_{u}^*(t),t),
\]
and evaluate $\bs A$ and $\bs B_j$
at $\bs z_{u}^*(t)$.

Equations~\eqref{eq:gradient_flow_identity}
and~\eqref{eq:hessian_flow_identity}
then become
\begin{equation}
\label{eq:gradient_flow_trajectory}
\bs g(t)
=
\bs A^\top\bs g(s),
\end{equation}
and
\begin{equation}
\label{eq:hessian_flow_trajectory}
\bs H(t)
=
\bs A^\top\bs H(s)\bs A
+
\sum_{j=1}^{2n}g_j(s)\bs B_j.
\end{equation}

\medskip
\noindent
\textbf{Step 3: Variational equations
of the characteristic flow.}

Differentiating the flow equation with
respect to its initial state gives
\begin{equation}
\label{eq:first_flow_variation}
\frac{\partial\bs A}{\partial s}
=
D_{\bs z}\bs b(\Phi_{t,s}(\bs z))
\bs A,
\qquad
\bs A(t,t,\bs z)=\bs I.
\end{equation}

Differentiating once more yields the
second-order variational equation
\begin{equation}
\label{eq:second_flow_variation}
\begin{aligned}
\frac{\partial\bs B_j}{\partial s}
={}&
\sum_{\ell=1}^{2n}
\frac{\partial b_j}{\partial z_\ell}
(\Phi_{t,s}(\bs z))
\bs B_\ell
+
\bs A^\top
D_{\bs z}^{2}b_j(\Phi_{t,s}(\bs z))
\bs A,
\end{aligned}
\end{equation}
with $\bs B_j(t,t,\bs z)=\bs 0.$

At $s=t$, we have
$\bs A(t,t,\bs z)=\bs I$ and
$\bs B_j(t,t,\bs z)=\bs 0$.
Therefore,
\[
\left.
\frac{\partial\bs A}{\partial s}
\right|_{s=t}
=
D_{\bs z}\bs b(\bs z) \text{ and } 
\left.
\frac{\partial\bs B_j}{\partial s}
\right|_{s=t}
=
D_{\bs z}^{2}b_j(\bs z).
\]

Taking $s=t+\varepsilon$ and evaluating
along $\bs z_{u}^*(t)$ gives
\begin{equation}
\label{eq:flow_A_expansion}
\bs A
=
\bs I+\varepsilon\bs J(t)
+o(\varepsilon),
\end{equation}
and
\begin{equation}
\label{eq:flow_B_expansion}
\bs B_j
=
\varepsilon
D_{\bs z}^{2}
F_j(\bs z^*(t),\bar u)
+o(\varepsilon),
\end{equation}
where $\bs J(t)
=
D_{\bs z}F(\bs z^*(t),\bar u).$

\medskip
\noindent
\textbf{Step 4: Evolution of the gradient.}

Setting $s=t+\varepsilon$ in
Eq.~\eqref{eq:gradient_flow_trajectory}
and using Eq.~\eqref{eq:flow_A_expansion},
we obtain
\[
\bs g(t)
=
\bigl(
\bs I+\varepsilon\bs J(t)^\top
+o(\varepsilon)
\bigr)
\bs g(t+\varepsilon).
\]

Since $\bs g$ is continuous locally,
\[
\begin{aligned}
\frac{\bs g(t+\varepsilon)-\bs g(t)}
{\varepsilon}
=
-\bs J(t)^\top\bs g(t)
+o(1).
\end{aligned}
\]

Due to the limit
$\varepsilon\to0,$ we can have
\begin{equation}
\label{eq:gradient_local_evolution}
\dot{\bs g}(t)
=
-\bs J(t)^\top\bs g(t).
\end{equation}

\medskip
\noindent
\textbf{Step 5: Evolution of the Hessian.}

Taking $s=t+\varepsilon$ in
Eq.~\eqref{eq:hessian_flow_trajectory}
gives
\[
\begin{aligned}
\bs H(t)
={}&
\bs A^\top
\bs H(t+\varepsilon)
\bs A
\\
&+
\sum_{j=1}^{d}
g_j(t+\varepsilon)\bs B_j.
\end{aligned}
\]

Substituting
Eqs.~\eqref{eq:flow_A_expansion}
and~\eqref{eq:flow_B_expansion},
and using the local continuity of
$\bs g$ and $\bs H$, yields
\[
\begin{aligned}
\bs H(t)
={}&
\bs H(t+\varepsilon)
+
\varepsilon
\bs J(t)^\top\bs H(t)
+
\varepsilon
\bs H(t)\bs J(t)
\\
&+
\varepsilon
\sum_{j=1}^{d}
g_j(t)
D_{\bs z}^{2}
F_j(\bs z^*(t),\bar u)
+
o(\varepsilon).
\end{aligned}
\]

Define
\[
\bs\Delta(t)=\nabla_{\bs z}
\!\left[
\bs J(t)^\top
\right]
\nabla_{\bs z}V(\bs z_{u}^*(t),t)
=
\sum_{j=1}^{d}
g_j(t)
D_{\bs z}^{2}
F_j(\bs z^*(t),u^*(t)).
\]

Since $u^*(t)=\bar u$ on the
local smooth branch, rearranging gives
\[
\begin{aligned}
\frac{\bs H(t+\varepsilon)-\bs H(t)}
{\varepsilon}
={}&
-\bs H(t)\bs J(t)
-\bs J(t)^\top\bs H(t)
\\
&-\bs\Delta(t)+o(1).
\end{aligned}
\]

Letting $\varepsilon\to0$, we obtain
\begin{equation}
\label{eq:hessian_local_evolution}
\dot{\bs H}(t)
=
-\bs H(t)\bs J(t)
-\bs J(t)^\top\bs H(t)
-\bs\Delta(t).
\end{equation}

Thus, both
Eqs.~\eqref{eq:gradient_local_evolution}
and~\eqref{eq:hessian_local_evolution}
hold locally at every
$t\in(0,T)\setminus\Sigma$.

\medskip
\noindent
\textbf{Step 6: Evolution on the full
time interval.}

By Assumption~\ref{ass:hamilton_maximizer},
the switching-time set $\Sigma$ has
Lebesgue measure zero.
Here, the values of $\bs J(t)$ and
$\bs\Delta(t)$ on the null set $\Sigma$
do not affect the almost-everywhere
identities.
Furthermore, $\bs g(t)$ and $\bs H(t)$
admit absolutely continuous extensions
to $[0,T]$.
Thus, Eq. \eqref{eq:gradient_local_evolution} and \eqref{eq:hessian_local_evolution} 
hold for almost
every $t\in[0,T]$ along the HJ-optimal trajectory.
\end{proof}

\paragraph{Proof of Theorem~\ref{thm:second_order_consistency}:}
\begin{proof}
Define the error matrix
\[
\bs E(t)=\bs P(t)-\bs H(t).
\]
Using equation~\eqref{eq:hj_continuation} and~\eqref{eq:P_evolution}, we obtain
\[
\dot{\bs E}(t)
=
-\bs E(t)\bs J(t)
-
\bs J(t)^\top\bs E(t)
-
\big(\bs\Delta_P(t)-\bs\Delta(t)\big).
\]
Since $\bs\Delta_P(t)=\bs\Delta(t)$ for almost every $t$, this reduces to
\[
\dot{\bs E}(t)
=
-\bs E(t)\bs J(t)
-
\bs J(t)^\top\bs E(t).
\]
The terminal condition $\bs P(T)=\bs H(T)$ implies
\[
\bs E(T)=0.
\]
By uniqueness of solutions to this linear matrix differential equation, we have
\[
\bs E(t)=0,
\qquad
\text{for a.e. } t\in[0,T].
\]
Therefore,
\[
\bs P(t)=\bs H(t),
\qquad
\text{for a.e. } t\in[0,T],
\]
which proves the result.
\end{proof}

\paragraph{Proof of Theorem~\ref{thm:first_order_consistency}:}
\begin{proof}
Define the Jacobian error
\[
\bs e(t)
=
\bs h_J(t)-\bs g(t).
\]
Using equation~\eqref{eq:value_jacobian_evolution}
and~\eqref{eq:gebf_jacobian_evolution}, we obtain
\[
\dot{\bs e}(t)
=
-\bs J(t)^\top \bs e(t),
\qquad
\text{for a.e. }t\in[0,T].
\]
The shared terminal condition gives
\[
\bs e(T)
=
\bs h_J(T)-\bs g(T)
=
0.
\]
By uniqueness of solutions to this linear terminal-value problem,
\[
\bs e(t)=0,
\qquad
\text{for a.e. }t\in[0,T].
\]
Therefore,
\[
\bs h_J(t)=\bs g(t),
\]
which proves
\[
\nabla_{\bs z}h(\bs z_{u}^*(t),t)
=
\nabla_{\bs z}V(\bs z_{u}^*(t),t)
\]
for almost every $t\in[0,T]$.
\end{proof}

\paragraph{Proof of Theorem~\ref{thm:zeroth_order_consistency}:}
\begin{proof}
Define the value error
\[
e(t)
=
h(\bs z_u^*(t),t)
-
V(\bs z_u^*(t),t).
\]
Using the chain rule, we have
\begin{align}
\dot e(t)
=&
\left[
\frac{\partial h}{\partial t}
+
\nabla_{\bs z}h^\top
F(\bs z_u^*(t),u^*(t))
\right]
\nonumber\\
&-
\left[
\frac{\partial V}{\partial t}
+
\nabla_{\bs z}V^\top
F(\bs z_u^*(t),u^*(t))
\right].
\end{align}
By construction, Eq.~\eqref{eq:h_value_evolution} implies
\[
\frac{\partial h}{\partial t}
+
\nabla_{\bs z}h^\top
F(\bs z_u^*(t),u^*(t))
=
0.
\]
Moreover, along the Hamiltonian-maximizing trajectory, the HJ equation gives
\[
\frac{\partial V}{\partial t}
+
\nabla_{\bs z}V^\top
F(\bs z_u^*(t),u^*(t))
=
0
\]
for almost every $t\in[0,T]$.
Therefore,
\[
\dot e(t)=0,
\qquad
\text{for a.e. }t\in[0,T].
\]
The terminal condition gives
\[
e(T)
=
h(\bs z_u^*(T),T)-V(\bs z_u^*(T),T)
=
l(\bs z_u^*(T))-l(\bs z_u^*(T))
=
0.
\]
Hence $e(t)=0$ for almost every $t\in[0,T]$, which proves
\[
h(\bs z_u^*(t),t)
=
V(\bs z_u^*(t),t).
\]
\end{proof}

\section{Supplementary of Section~\ref{sec:algorithm}}\label{append_sec:algorithm}

\paragraph{Proof of Theorem~\ref{thm:fixed_point_consistency}:}
\begin{proof}
Let the reference trajectory supplied to
Algorithm~\ref{alg:GEBF} be the HJ-optimal trajectory,
\begin{equation*}
    \bs z_{\mathrm{ref}}(t)=\bs z_{u}^*(t).
\end{equation*}

Define the Hamiltonian-maximizing control by
\begin{equation}
\label{eq:maximizer_map}
    \mu(\bs z,\bs p)
    \in
    \arg\max_{u\in\mathcal U}
    \bs p^\top\bs F(\bs z,u),
\end{equation}
which is single-valued almost everywhere along the characteristic
under the uniqueness assumption in Assumption \ref{ass:hamilton_maximizer}.

Let
\begin{equation}
\label{eq:true_costate}
    \bs g^*(t)
    :=
    \nabla_{\bs z}V(\bs z_{u}^*(t),t)
\end{equation}
at the times at which $V$ is differentiable along $\bs z_{u}^*(t)$.
By the characteristic equations associated with the HJ PDE,
$\bs g^*$ satisfies
\begin{equation}
\label{eq:true_costate_dynamics}
    \dot{\bs g}^*(t)
    =
    -
    \bs J^*(t)^\top \bs g^*(t)
\end{equation}
for almost every $t\in[0,T]$, where
\begin{equation*}
    \bs J^*(t)
    =
    \nabla_{\bs z}\bs F
    \bigl(
        \bs z_{u}^*(t),u^*(t)
    \bigr)
\end{equation*}
and
\begin{equation}
\label{eq:true_optimal_control}
    u^*(t)
    =
    \mu\bigl(
        \bs z_{u}^*(t),\bs g^*(t)
    \bigr)
\end{equation}
almost everywhere.

Using \eqref{eq:true_optimal_control},
\eqref{eq:true_costate_dynamics} can equivalently be written as
\begin{equation}
\label{eq:true_costate_closed}
    \dot{\bs g}^*(t)
    =
    -
    \nabla_{\bs z}\bs F
    \left(
        \bs z_{u}^*(t),
        \mu(\bs z_{u}^*(t),\bs g^*(t))
    \right)^\top
    \bs g^*(t).
\end{equation}

Since $V(\bs z,T)=l(\bs z)$, the relevant terminal condition is
\begin{equation}
\label{eq:true_costate_terminal}
    \bs g^*(T)
    =
    \nabla_{\bs z}l(\bs z_{u}^*(T)).
\end{equation}

Now consider the GEBF constructed by
Algorithm~\ref{alg:GEBF} using $\bs z_{u}^*$ as its reference trajectory.
Let
\begin{equation*}
    h(\bs z,t)
    =
    \frac{1}{2}\bs z^\top\bs P(t)\bs z
    +
    \bs Q(t)^\top\bs z
    +
    c(t),
\end{equation*}
and define its gradient along the reference trajectory by
\begin{equation}
\label{eq:gebf_costate}
    \bs g_h(t)
    :=
    \nabla_{\bs z}h(\bs z_{u}^*(t),t)
    =
    \bs P(t)\bs z_{u}^*(t)+\bs Q(t).
\end{equation}

By the first-order GEBF propagation equation used in
Algorithm~\ref{alg:GEBF}, $\bs g_h$ satisfies
\begin{equation}
\label{eq:gebf_costate_dynamics}
    \dot{\bs g}_h(t)
    =
    -
    \bs J_h(t)^\top\bs g_h(t)
\end{equation}
for almost every $t\in[0,T]$, where
\begin{equation*}
    \bs J_h(t)
    =
    \nabla_{\bs z}\bs F
    \bigl(
        \bs z_{u}^*(t),u_h(t)
    \bigr),
\end{equation*}
and the control induced by the GEBF is
\begin{equation}
\label{eq:gebf_control}
    u_h(t)
    =
    \mu\bigl(
        \bs z_{u}^*(t),\bs g_h(t)
    \bigr).
\end{equation}
Consequently,
\eqref{eq:gebf_costate_dynamics} can be written as
\begin{equation}
\label{eq:gebf_costate_closed}
    \dot{\bs g}_h(t)
    =
    -
    \nabla_{\bs z}\bs F
    \left(
        \bs z_{u}^*(t),
        \mu(\bs z_{u}^*(t),\bs g_h(t))
    \right)^\top
    \bs g_h(t).
\end{equation}

The terminal GEBF geometry is initialized from the terminal function
$l$. Hence,
\begin{equation}
\label{eq:gebf_costate_terminal}
    \bs g_h(T)
    =
    \nabla_{\bs z}l(\bs z_{u}^*(T)).
\end{equation}

Equations \eqref{eq:true_costate_closed} and
\eqref{eq:gebf_costate_closed} therefore define the same backward
dynamics, and
\eqref{eq:true_costate_terminal} and
\eqref{eq:gebf_costate_terminal} give the same terminal condition.
Based on Theorem \ref{thm:first_order_consistency},
\begin{equation}
\label{eq:gradient_consistency_fixed}
    \bs g_h(t)
    =
    \bs g^*(t)
    =
    \nabla_{\bs z}V(\bs z_{u}^*(t),t)
\end{equation}
for almost every $t\in[0,T]$.

It follows from \eqref{eq:gebf_control},
\eqref{eq:true_optimal_control}, and
\eqref{eq:gradient_consistency_fixed} that
\begin{equation}
\label{eq:control_consistency_fixed}
    u_h(t)
    =
    u^*(t)
\end{equation}
for almost every $t\in[0,T]$.

We finally consider the forward trajectory update in
Algorithm~\ref{alg:GEBF}. Let $\widehat{\bs z}(t)$ denote the
trajectory generated by the GEBF-induced control. It satisfies
\begin{equation}
\label{eq:updated_trajectory}
    \dot{\widehat{\bs z}}(t)
    =
    \bs F(\widehat{\bs z}(t),u_h(t)),
    \qquad
    \widehat{\bs z}(0)=\bs z_0.
\end{equation}
Using \eqref{eq:control_consistency_fixed},
\begin{equation*}
    \dot{\widehat{\bs z}}(t)
    =
    \bs F(\widehat{\bs z}(t),u^*(t))
\end{equation*}
for almost every $t\in[0,T]$.

On the other hand, the HJ-optimal trajectory satisfies
\begin{equation*}
    \dot{\bs z}_{u}^*(t)
    =
    \bs F(\bs z_{u}^*(t),u^*(t)),
    \qquad
    \bs z_{u}^*(0)=\bs z_0.
\end{equation*}
Thus $\widehat{\bs z}$ and $\bs z_{u}^*$ solve the same state equation
under the same control and with the same initial condition.
By uniqueness of the forward state trajectory,
\begin{equation*}
    \widehat{\bs z}(t)
    =
    \bs z_{u}^*(t),
    \qquad
    t\in[0,T].
\end{equation*}

Therefore, applying one trajectory-update step of
Algorithm~\ref{alg:GEBF} to the HJ-optimal trajectory leaves the
trajectory unchanged:
\begin{equation*}
    \mathcal T(\bs z_{u}^*)=\bs z_{u}^*.
\end{equation*}
Hence $\bs z_{u}^*$ is a fixed point of the GEBF trajectory iteration.
\end{proof}

\paragraph{Proof of Theorem~\ref{thm:trajectory_lower_bound}:}
\begin{proof}
Define the Hamiltonian by
\begin{equation*}
    H(\bs z,\bs p)
    :=
    \max_{u\in\mathcal U}
    \bs p^\top \bs F(\bs z,u).
\end{equation*}
Since
\begin{equation*}
    u_{\mathrm{ref}}(t)
    \in
    \arg\max_{u\in\mathcal U}
    \nabla_{\bs z}h(\bs z_{\mathrm{ref}}(t),t)^\top
    \bs F(\bs z_{\mathrm{ref}}(t),u),
\end{equation*}
we have
\begin{equation}
\label{eq:Href}
    H\!\left(
        \bs z_{\mathrm{ref}}(t),
        \nabla_{\bs z}h(\bs z_{\mathrm{ref}}(t),t)
    \right)
    =
    \nabla_{\bs z}h(\bs z_{\mathrm{ref}}(t),t)^\top
    \bs F(\bs z_{\mathrm{ref}}(t),u_{\mathrm{ref}}(t)).
\end{equation}

By the assumed GEBF residual condition,
\begin{equation}
\label{eq:h_residual_proof}
   \frac{\partial h}{\partial t}(\bs z_{\mathrm{ref}}(t),t)
    +
    H\!\left(
        \bs z_{\mathrm{ref}}(t),
        \nabla_{\bs z}h(\bs z_{\mathrm{ref}}(t),t)
    \right)
    =0
\end{equation}
for almost every $t\in[0,T]$.
Combining \eqref{eq:Href} and \eqref{eq:h_residual_proof} yields
\begin{equation}
\label{eq:ht_ref}
    \frac{\partial h}{\partial t}(\bs z_{\mathrm{ref}}(t),t)
    =
    -
    \nabla_{\bs z}h(\bs z_{\mathrm{ref}}(t),t)^\top
    \bs F(\bs z_{\mathrm{ref}}(t),u_{\mathrm{ref}}(t))
\end{equation}
for almost every $t\in[0,T]$.

Since the reference trajectory satisfies $\dot{\bs z}_{\mathrm{ref}}(t)
    =
    \bs F(\bs z_{\mathrm{ref}}(t),u_{\mathrm{ref}}(t))$
for almost every $t\in[0,T]$, the chain rule gives
\begin{align}
    \frac{d}{dt}
    h(\bs z_{\mathrm{ref}}(t),t)
    &=
    \frac{\partial h}{\partial t}(\bs z_{\mathrm{ref}}(t),t)
    +
    \nabla_{\bs z}h(\bs z_{\mathrm{ref}}(t),t)^\top
    \dot{\bs z}_{\mathrm{ref}}(t)
    \nonumber\\
    &=
   \frac{\partial h}{\partial t}(\bs z_{\mathrm{ref}}(t),t)
    +
    \nabla_{\bs z}h(\bs z_{\mathrm{ref}}(t),t)^\top
    \bs F(\bs z_{\mathrm{ref}}(t),u_{\mathrm{ref}}(t))
    \nonumber\\
    &=0
\label{eq:h_constant_ref}
\end{align}
for almost every $t\in[0,T]$.

We next consider the HJ value function along the same reference
trajectory. At almost every $t$ for which $V$ is differentiable along
$\bs z_{\mathrm{ref}}(t)$, the chain rule gives
\begin{align}
    \frac{d}{dt}
    V(\bs z_{\mathrm{ref}}(t),t)
    &=
    \frac{\partial V}{\partial t}(\bs z_{\mathrm{ref}}(t),t)
    +
    \nabla_{\bs z}V(\bs z_{\mathrm{ref}}(t),t)^\top
    \dot{\bs z}_{\mathrm{ref}}(t)
    \nonumber\\
    &=
    \frac{\partial V}{\partial t}(\bs z_{\mathrm{ref}}(t),t)
    +
    \nabla_{\bs z}V(\bs z_{\mathrm{ref}}(t),t)^\top
    \bs F(\bs z_{\mathrm{ref}}(t),u_{\mathrm{ref}}(t)).
\label{eq:V_chain_ref}
\end{align}

Since $V$ satisfies the HJ equation $\frac{\partial V}{\partial t}(\bs z,t)
    +
    H\!\left(
        \bs z,\nabla_{\bs z}V(\bs z,t)
    \right)
    =0,$
we have, at such times,
\begin{equation*}
    \frac{\partial V}{\partial t}(\bs z_{\mathrm{ref}}(t),t)
    =
    -
    H\!\left(
        \bs z_{\mathrm{ref}}(t),
        \nabla_{\bs z}V(\bs z_{\mathrm{ref}}(t),t)
    \right).
\end{equation*}
Substituting this relation into \eqref{eq:V_chain_ref} gives
\begin{align}
    \frac{d}{dt}
    V(\bs z_{\mathrm{ref}}(t),t)
    &=
    -
    H\!\left(
        \bs z_{\mathrm{ref}}(t),
        \nabla_{\bs z}V(\bs z_{\mathrm{ref}}(t),t)
    \right)
    \nonumber\\
    &\quad+
    \nabla_{\bs z}V(\bs z_{\mathrm{ref}}(t),t)^\top
    \bs F(\bs z_{\mathrm{ref}}(t),u_{\mathrm{ref}}(t))
    \nonumber\\
    &=
    -\delta_H(t)
\label{eq:Vdot_delta}
\end{align}
for almost every $t\in[0,T]$.

By the definition of the Hamiltonian,
\begin{equation*}
\begin{aligned}
    H\!\left(
        \bs z_{\mathrm{ref}}(t),
        \nabla_{\bs z}V(\bs z_{\mathrm{ref}}(t),t)
    \right)
    =
    \max_{u\in\mathcal U}
    \nabla_{\bs z}V(\bs z_{\mathrm{ref}}(t),t)^\top
    \bs F(\bs z_{\mathrm{ref}}(t),u).
\end{aligned}
\end{equation*}
Therefore, for the particular admissible control
$u_{\mathrm{ref}}(t)\in\mathcal U$,
\begin{equation*}
\begin{aligned}
    H\!\left(
        \bs z_{\mathrm{ref}}(t),
        \nabla_{\bs z}V(\bs z_{\mathrm{ref}}(t),t)
    \right)
    \ge
    \nabla_{\bs z}V(\bs z_{\mathrm{ref}}(t),t)^\top
    \bs F(\bs z_{\mathrm{ref}}(t),u_{\mathrm{ref}}(t)).
\end{aligned}
\end{equation*}
Consequently, $\delta_H(t)\ge0$
for almost every $t\in[0,T]$.

Now define the trajectory-wise approximation error $e(t)
    :=
    V(\bs z_{\mathrm{ref}}(t),t)
    -
    h(\bs z_{\mathrm{ref}}(t),t).$
Combining \eqref{eq:h_constant_ref} and
\eqref{eq:Vdot_delta}, we obtain
\begin{equation}
\label{eq:error_dynamics}
    \dot e(t)
    =
    -\delta_H(t)
\end{equation}
for almost every $t\in[0,T]$.

Since $e$ is absolutely continuous, integrating
\eqref{eq:error_dynamics} from $t$ to $T$ gives $ e(T)-e(t)
    =
    -\int_t^T \delta_H(s)\,ds.$
The terminal matching condition implies $e(T)
    =
    V(\bs z_{\mathrm{ref}}(T),T)
    -
    h(\bs z_{\mathrm{ref}}(T),T)
    =0.$
Hence,
\begin{equation}
\label{eq:gap_identity}
    V(\bs z_{\mathrm{ref}}(t),t)
    -
    h(\bs z_{\mathrm{ref}}(t),t)
    =
    \int_t^T \delta_H(s)\,ds.
\end{equation}

Since $\delta_H(s)\ge0$ for almost every $s\in[t,T]$,
\begin{equation*}
    \int_t^T \delta_H(s)\,ds\ge0.
\end{equation*}
It therefore follows from \eqref{eq:gap_identity} that
\begin{equation*}
    h(\bs z_{\mathrm{ref}}(t),t)
    \le
    V(\bs z_{\mathrm{ref}}(t),t),
    \qquad
    t\in[0,T].
\end{equation*}
This completes the proof.
\end{proof}

\section{Supplementary of Section~\ref{sec:experiments}}\label{append_experiments}

\subsection{Supplementary of Example.~\ref{append_experiments_example:one}}\label{append_append_experiments_example:one}

\begin{figure}[htb!]
    \centering
    \includegraphics[width=\linewidth]{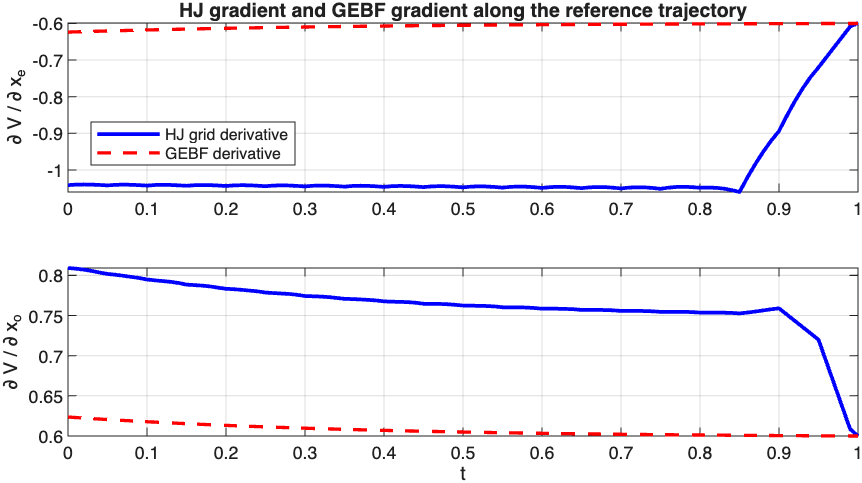}
    \caption{\textcolor{black}{Comparison of the gradient components of the
GEBF $h$ and the HJ value function $V$.}
    }
    \label{fig:case1_1D_gradient_ac}
\end{figure}

\begin{figure}[htb!]
    \centering
    \includegraphics[width=\linewidth]{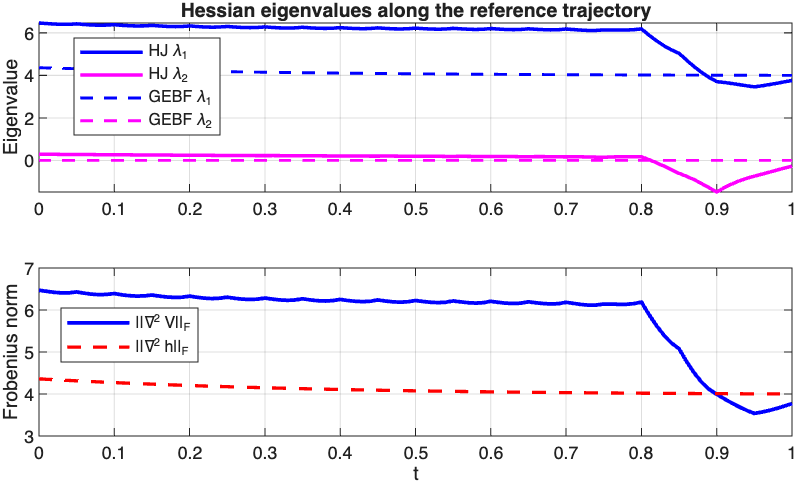}
    \caption{\textcolor{black}{Comparison of the Hessian components of the
GEBF $h$ and the HJ value function $V$.}
    }
    \label{fig:case1_1D_Hessian_ac}
\end{figure}

\textcolor{black}{Figures 8 and 9 illustrate the numerical evolution of the HJ gradient and Hessian along the reference trajectory. The gradient components, Hessian eigenvalues, and Frobenius norm exhibit no apparent jump discontinuities, despite more pronounced variations near the terminal time. These observations are consistent with the absolute continuity assumptions on the first- and second-order HJ derivatives in Assumption~\ref{ass:hj_second_order}.
}

\subsection{Supplementary of Example.~\ref{example:three_dimension}}\label{append_experiments_example:three}

\paragraph{Model.}
The ego vehicle follows the constant-speed kinematic car dynamics, a standard benchmark in helperOC~\cite{mitchell2005toolbox,fisac2015reach} and autonomous driving~\cite{yu2023sequential,chen2025int2planner}
\begin{equation}
\begin{aligned}
    \dot p^e_x = v_e \cos \theta_e,~
    \dot p^e_y = v_e \sin \theta_e,~ \dot \theta_e = u,
\end{aligned}
\label{eq:ego_unicycle}
\end{equation}
where $p^e=(p^e_x,p^e_y)$ is the position, $\theta_e$ is its heading angle, and $ u\in\mathcal U=[-\bar u,\bar u]$. The obstacle moves with constant speed and fixed heading,
\begin{equation}
\begin{aligned}
    \dot p^o_x = v_o \cos \theta_o,~\dot p^o_y = v_o \sin \theta_o,~
    \dot \theta_o=0 .
\end{aligned}
\label{eq:obs_unicycle}
\end{equation}

\paragraph{Derivation of the relative dynamics of Example.~\ref{example:three_dimension}.}
Let $x_1 =
    \cos\theta_e (p^o_x-p^e_x)
    +
    \sin\theta_e (p^o_y-p^e_y),$ 
    $x_2 =
    -\sin\theta_e (p^o_x-p^e_x)
    +
    \cos\theta_e (p^o_y-p^e_y),$ and $\theta = \theta_o-\theta_e.$
Here $[x_1,x_2]$ denotes the obstacle position in the ego body frame, and $\theta$ is relative heading angle.
Let
\[
r=p^o-p^e,
\qquad
R_e=
\begin{bmatrix}
\cos\theta_e & \sin\theta_e\\
-\sin\theta_e & \cos\theta_e
\end{bmatrix},
\qquad
\begin{bmatrix}x_1\\x_2\end{bmatrix}
=
R_e r .
\]
Using
\[
\dot r =
\begin{bmatrix}
v_o\cos\theta_o-v_e\cos\theta_e\\
v_o\sin\theta_o-v_e\sin\theta_e
\end{bmatrix}
\]
and $\dot\theta_e=u$, direct differentiation gives
\[
\dot x_1=-v_e+v_o\cos(\theta_o-\theta_e)+u x_2,
\]
\[
\dot x_2=v_o\sin(\theta_o-\theta_e)-u x_1,
\]
and
\[
\dot\theta=\dot\theta_o-\dot\theta_e=-u.
\]
Thus, with $\theta=\theta_o-\theta_e$, we obtain 
\begin{equation}
\begin{aligned}
    \dot x_1 = -v_e + v_o\cos\theta + u x_2,~
    \dot x_2 =  v_o\sin\theta - u x_1,~
    \dot \theta &= -u .
\end{aligned}
\label{eq:air3d_nodisturbance}
\end{equation}

\begin{figure}
    \centering
    \includegraphics[width=\linewidth]{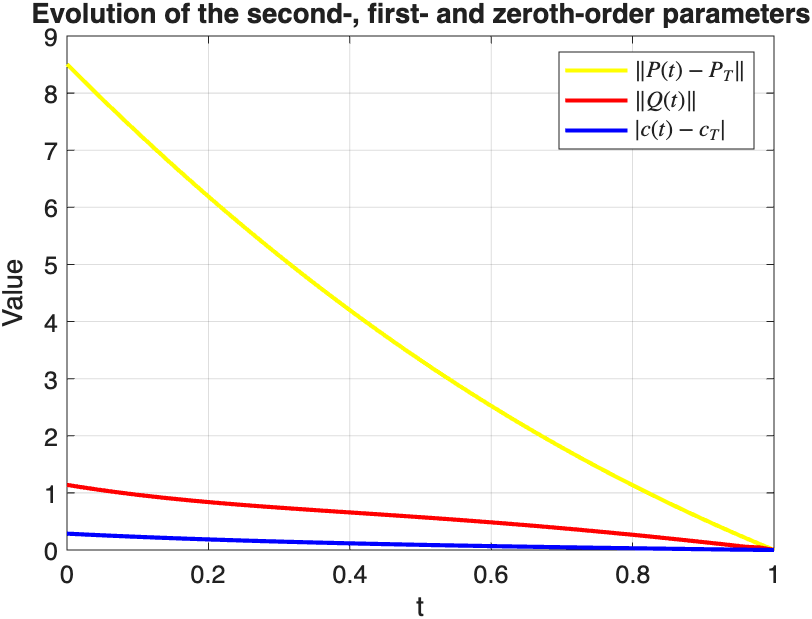}
    \caption{Evolution of GEBF parameters.
    }
    \label{fig:case2_parameters_evolution}
\end{figure}

\begin{figure}[htb!]
    \centering
    \includegraphics[width=\linewidth]{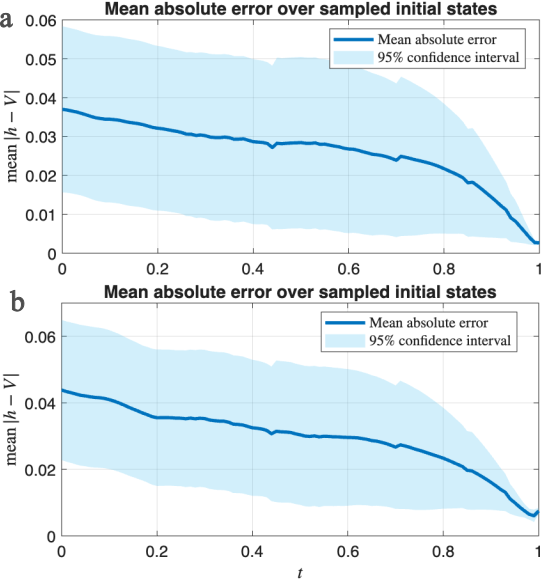}
    \caption{(a) Mean absolute error over sampled initial state under the unform grid of size $121\times 121\times 101$.
    (b) Mean absolute error over sampled initial state under the unform grid of size $81\times 81\times 81$.}
    \label{fig:mean_sample_comparison}
\end{figure}

\begin{figure}
    \centering
    \includegraphics[width=\linewidth]{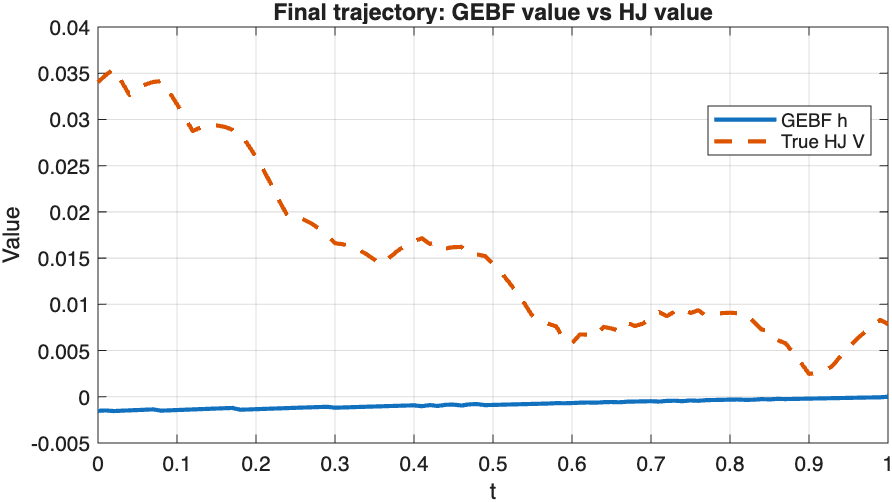}
    \caption{\textcolor{black}{GEBF and HJ value comparison near the zero level set.}
    }
    \label{fig:case2_value_comparsion_levelset}
\end{figure}

\begin{figure}[htb!]
    \centering
    \includegraphics[width=\linewidth]{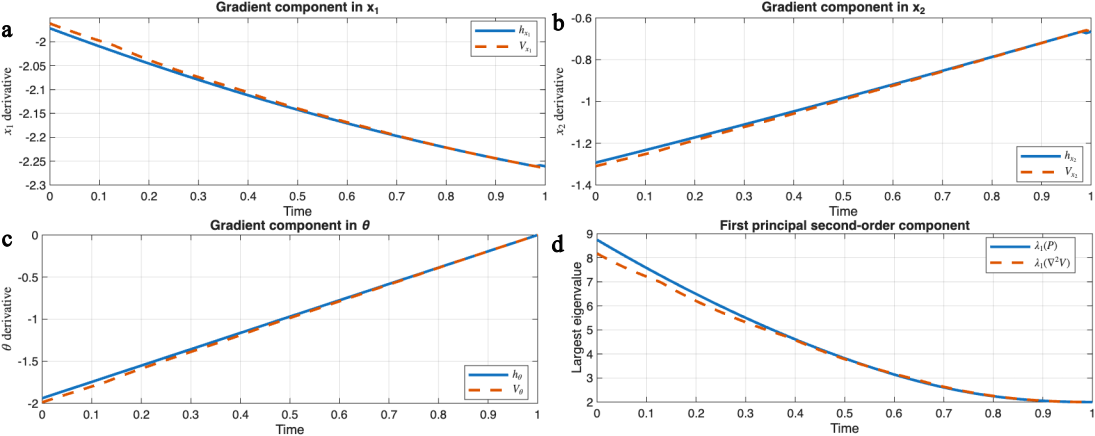}
    \caption{\textcolor{black}{(a)--(c) Comparison of the gradient components of the
GEBF $h$ and the HJ value function $V$.
(d) Comparison of the largest eigenvalues of
$\nabla_{\bs z}^2 h$ and $\nabla_{\bs z}^2 V$.}
    }
    \label{fig:case2_ac_validation}
\end{figure}

\paragraph{Parameters Setting.}
In the experiment, we use $ v_e=1,~v_o=1.5,~\bar u=0.3$ and $ R=1.$
The computational domain is $x_1,x_2\in[-6,6]$ and $\theta\in[-\pi,\pi].$
We use a uniform grid of size
$121\times121\times101$ and time step $\Delta t=0.01$ over the horizon $T=1$.

\subsection{Supplementary of Example.~\ref{ex:8d_simple_car}}\label{append_ex:8d_simple_car}

\paragraph{Model.}
The ego dynamics are 
\begin{equation}
\begin{aligned}
    \dot p_x^e = v_e\cos\theta_e,~
    \dot p_y^e = v_e\sin\theta_e,~
    \dot\theta_e = \omega_e,~
    \dot v_e = a_e,
\end{aligned}
\label{eq:8d_ego_dynamics}
\end{equation}
where $u=[a_e , \omega_e]^{\top}
    \in\mathcal U$
is the control input, consisting of the longitudinal acceleration
$a_e\in [-1.5,1.5]$ and angular velocity $\omega_e\in [-1,1]$.
The obstacle dynamics are
\begin{equation}
\begin{aligned}
    \dot p_x^o = v_o\cos\theta_o,~
    \dot p_y^o = v_o\sin\theta_o,~
    \dot\theta_o = \omega_o,~
    \dot v_o = a_o,
\end{aligned}
\label{eq:8d_obstacle_dynamics}
\end{equation}
where  $a_o$ and $\omega_o$ are the prescribed obstacle acceleration and angular velocity, respectively.

\paragraph{5D Form.}
We reduce the system to 5D relative dynamics using the identical technique described in the previous example
\begin{equation}
\begin{aligned}
    \dot x_1
    &=
    v_o\cos\psi-v_e+\omega_e x_2,~
    \dot x_2
    =
    v_o\sin\psi-\omega_e x_1,\\
    \dot\theta
    &=
    -\omega_e,~
    \dot v_e
    =
    a_e,~
    \dot v_o
    =
    a_o.
\end{aligned}
\label{eq:5d_relative_simple_car}
\end{equation}

\paragraph{Parameters Setting.}
In the experiment, 
the computational domain is $x_1,x_2\in[-6,6]$, $\theta\in[-\pi,\pi],$
and $v_{e},v_{o}\in [0,3],$ and collision radius is $R=1$. 
We use a uniform grid of size
$25\times 25\times 21\times 15\times 15$ and time step $\Delta t=0.01$ over the horizon $T=1.5$ for calculation of the HJ value function.

\subsection{Example: 8D Coupled System Without Relative-Coordinate Reduction}
\label{append_ex:8d_simple_car_noRela}

\textbf{Model.} 
We further consider an 8D planar collision-avoidance problem for which the standard relative-coordinate reduction is not applicable.
The system consists of a controlled ego vehicle and a non-adversarial moving obstacle, both modeled as 4D kinematic cars.
Their states are defined as
\[
\bs x_e
=
\begin{bmatrix}
p_x^e & p_y^e & \theta_e & v_e
\end{bmatrix}^{\top},
\qquad
\bs x_o
=
\begin{bmatrix}
p_x^o & p_y^o & \theta_o & v_o
\end{bmatrix}^{\top},
\]
where $(p_x^e,p_y^e)$ and $(p_x^o,p_y^o)$ denote the positions of the ego vehicle and obstacle, $\theta_e$ and $\theta_o$ are their heading angles, and $v_e$ and $v_o$ are their speeds.

The ego vehicle follows
\begin{equation}
\begin{aligned}
    \dot p_x^e &= v_e\cos\theta_e,\\
    \dot p_y^e &= v_e\sin\theta_e,\\
    \dot\theta_e &= \omega_e,\\
    \dot v_e &= a_e,
\end{aligned}
\label{eq:8d_coupled_ego}
\end{equation}
with control input
\[
u=
\begin{bmatrix}
a_e & \omega_e
\end{bmatrix}^{\top}
\in\mathcal U,
\]
where $a_e\in [-1,1]$ and $\omega_e\in[-0.6,0.6]$ denote the ego vehicle's acceleration and angular velocity, respectively.

The obstacle dynamics are given by
\begin{equation}
\begin{aligned}
    \dot p_x^o &= v_o\cos\theta_o,\\
    \dot p_y^o &= v_o\sin\theta_o,\\
    \dot\theta_o &= \omega_o,\\
    \dot v_o &= \alpha\tanh(p_x^o+1),
\end{aligned}
\label{eq:8d_coupled_obstacle}
\end{equation}
where $\omega_o$ is prescribed and $\alpha>0$ determines the dependence of the obstacle acceleration on its absolute position.
We build up the coupled dynamics in form of \eqref{eq:coupledDynamics}, where $\bs z=(\bs x_e-\bs x_o,\bs x_o)$.
Define $x_1:=p^e_x-p^o_x$ and $x_2:=p^e_y-p^o_y.$
The terminal function is $l(\bs z)=x_1^2+x_2^2-R^2.$

\paragraph{Parameters Setting.}
In the experiment, 
the computational domain is $x_1,x_2\in[-6,6]$, $\theta_{e},\theta_o\in[-\pi,\pi],$
and $v_{e},v_{o}\in [0,3],$ and collision radius is $R=0.5$.
We fix $\omega_o=0$ and $\alpha=0.8.$

\begin{figure}[htb!]
    \centering
    \includegraphics[width=\linewidth]{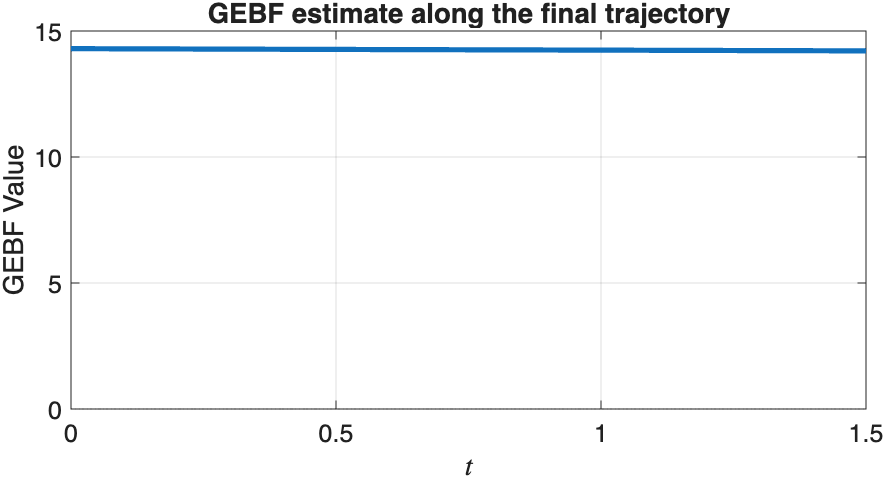}
    \caption{GEBF value}
    \label{fig:8D_withoutR_GEBF_value}
\end{figure}

\begin{figure}[htb!]
    \centering
    \includegraphics[width=\linewidth]{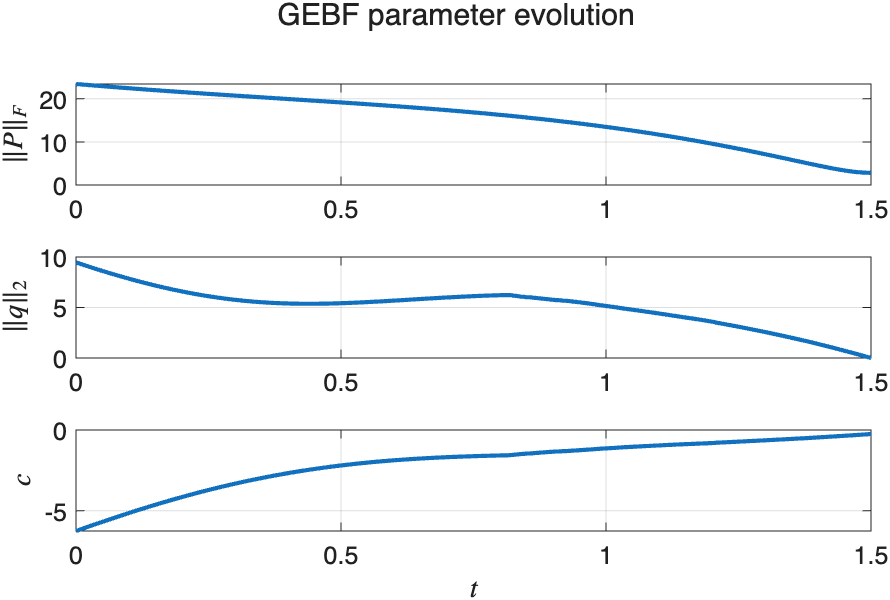}
    \caption{GEBF parameters eovlution}
    \label{fig:8D_withoutR_GEBF_parameter}
\end{figure}

Fig.~\ref{fig:8D_withoutR_GEBF_value} shows the evolution of the GEBF value along the final trajectory. Consistent with the HJ value function, the GEBF value remains nearly constant over time, reflecting the HJ equation
$\frac{\partial V}{\partial t}
\!+\!
\max_{u \in \mathcal U}
\nabla_{\bs z} V \!\cdot\! F(\bs z,u)\!
=0$.
Fig.~\ref{fig:8D_withoutR_GEBF_parameter} shows the continuous evolution of the GEBF parameters in~\eqref{eq:gebf}.

\begin{figure}[htb!]
    \centering
    \includegraphics[width=0.8\linewidth]{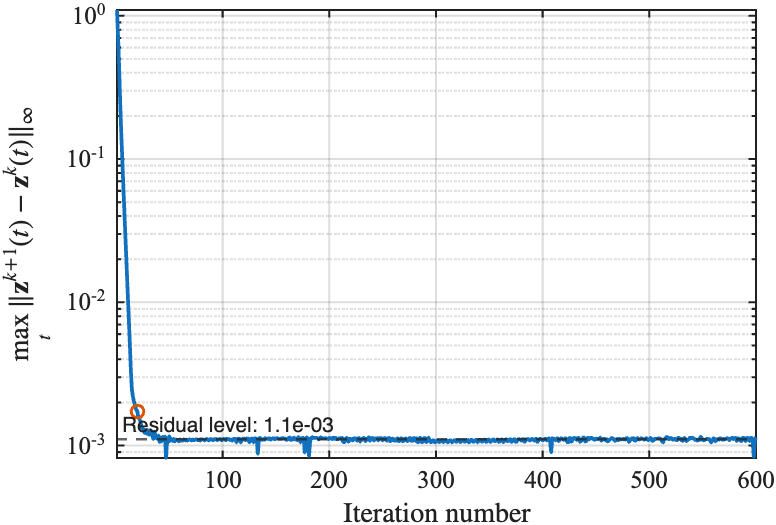}
    \caption{Trajectory-update error
over the GEBF refinement iterations.}
    \label{fig:error_iteration}
\end{figure}

\paragraph{Numerical convergence behaviour.}
Figure~\ref{fig:error_iteration} reports the trajectory-update
error
\[
e_k=
\max_{t\in[0,T]}
\left\|
\bs z^{k+1}(t)-\bs z^k(t)
\right\|_\infty
\]
over the refinement iterations.
The error decreases rapidly from order one to order \(10^{-3}\) within
approximately 20 iterations and subsequently remains at a small residual
level.
This behaviour indicates that the iterative trajectory refinement rapidly
approaches a numerically stable neighbourhood.

\end{document}